\documentclass[sigconf]{acmart}
\usepackage{appendix}
\usepackage{booktabs}
\usepackage{balance}
\usepackage{comment}
\usepackage{graphics}
\usepackage{verbatim}
\usepackage{epsfig}
\usepackage{latexsym}
\usepackage{subfigure}
\usepackage{algorithm}
\usepackage{color}
\usepackage{amsmath}
\usepackage{tabularx}
\usepackage{multirow}
\usepackage[bottom]{footmisc}
\usepackage{algpseudocode}
\usepackage{amssymb}
\usepackage[english]{babel}
\usepackage{balance}
\usepackage{blindtext}
\usepackage{caption}
\usepackage{enumitem}
\usepackage{url}
\usepackage{hhline}
\usepackage{times}
\usepackage[section]{placeins}
\usepackage{float}
\usepackage{multicol}
\usepackage{breakurl}

\setcopyright{none}
\renewcommand\footnotetextcopyrightpermission[1]{} 

\definecolor{red}{rgb}{0,0,0}
\begin{document}
	
	\title{WearID: Wearable-Assisted Low-Effort Authentication to Voice Assistants using Cross-Domain Speech Similarity}

	\author{Chen Wang}
	\orcid{0001-9737-1673}
	\affiliation{%
		\institution{Louisiana State University}
		\city{Baton Rouge}
		\state{LA}
		\country{USA}
		\postcode{70803}
			\thanks{This work was done during Chen Wang's Ph.D. study at Rutgers University.}
	}
	\email{chenwang1@lsu.edu}

	\author{Cong Shi}
	\affiliation{%
		\institution{WINLAB, Rutgers University}
		\city{New Brunswick}
		\state{NJ}
		\country{USA}
		\postcode{08901}
	}
	\email{cs1421@scarletmail.rutgers.edu}
	
		\author{Yingying Chen}
	\affiliation{%
		\institution{WINLAB, Rutgers University}
		\city{New Brunswick}
		\state{NJ}
		\country{USA}
		\postcode{08901}
	}
	\email{yingche@scarletmail.rutgers.edu}
	
	\author{Yan Wang}
	\affiliation{%
		\institution{Binghamton University}
		\city{Binghamton}
		\state{NY}
		\country{USA}
		\postcode{13902}
	}
	\email{yanwang@binghamton.edu}
	
	\author{Nitesh Saxena}
	\affiliation{%
		\institution{University of Alabama at Birmingham}
		\city{Birmingham}
		\state{AL}
		\country{USA}
		\postcode{35294}
	}
	\email{saxena@uab.edu}

	\begin{abstract}

Due to the open nature of voice input, voice assistant (VA) systems (e.g.,	Google Home and Amazon Alexa) are under a high risk of sensitive information leakage (e.g., personal schedules and shopping accounts). Though the existing VA systems may employ voice features to identify users, they are still vulnerable to various acoustic attacks	(e.g., impersonation, replay and hidden command attacks). In this work, we focus on the security issues of the emerging VA systems and aim to protect the users' highly sensitive information from these attacks. Towards this end, we propose a system, \textit{WearID}, which uses an off-the-shelf wearable device (e.g., a smartwatch or bracelet) as a secure token to verify the user's voice commands to the VA system. In particular, WearID exploits the readily available motion sensors from most wearables to describe the command sound in vibration domain and check the received command sound across two domains (i.e., wearable's motion sensor vs. VA device's microphone) to ensure the sound is from the legitimate user.

The \textit{cross-domain} design (audio vs. vibration) of our system leverages the motion sensor's shorter response distance to sounds (e.g., 25 cm), distinct sensing interface and its wide availability on wearable devices to secure the voice access, even when the microphone has been compromised in various acoustic attacks. However, examining the similarity of two sensing modalities is not trivial. The huge sampling rate gap (e.g., 8000Hz vs. 200Hz) causes the two data types hard to compare and even tiny data noises are magnified during such comparison. Moreover, as not designed for capturing sounds, the motion sensors show distinct response characteristics to sounds in terms of amplitude and frequency. In this work, we investigate the complex relationship between the two sensing modalities and develop an algorithm to convert the microphone data into low-frequency data comparable to the acoustic responses on the motion sensor. Our system further examines the similarity of the command sounds described in two domains to verify whether the voice command originates from the legitimate user. We report on extensive experiments to evaluate the WearID system under various audible and inaudible attacks. The results show that our system can verify the voice commands with $99.8\%$ accuracy in the normal situation and detect $97\%$ fake voice commands from the various impersonation and replay attacks and hidden voice and ultrasound attacks.
\end{abstract}

	\maketitle
	
	\vspace{-2mm}
\section{Introduction}

In recent years, smart devices (e.g., Google Home, Amazon Alexa, and smartphones) have incorporated advanced speech recognition technologies that enable the devices to understand natural language and take voice commands. By using voices as inputs, users can smoothly and conveniently interact with their voice assistant (VA) systems to accomplish numerous daily tasks. In particular, such a convenient function has been quickly adopted by users and widely used in various applications (e.g., playing music, managing calendar events, shopping online and controlling smart home appliances). As a result, VA systems have already been widely used in various scenarios, such as home, workplace and even public places.

 \begin{figure}[t]
 	\centering
 	\includegraphics[width=3in]{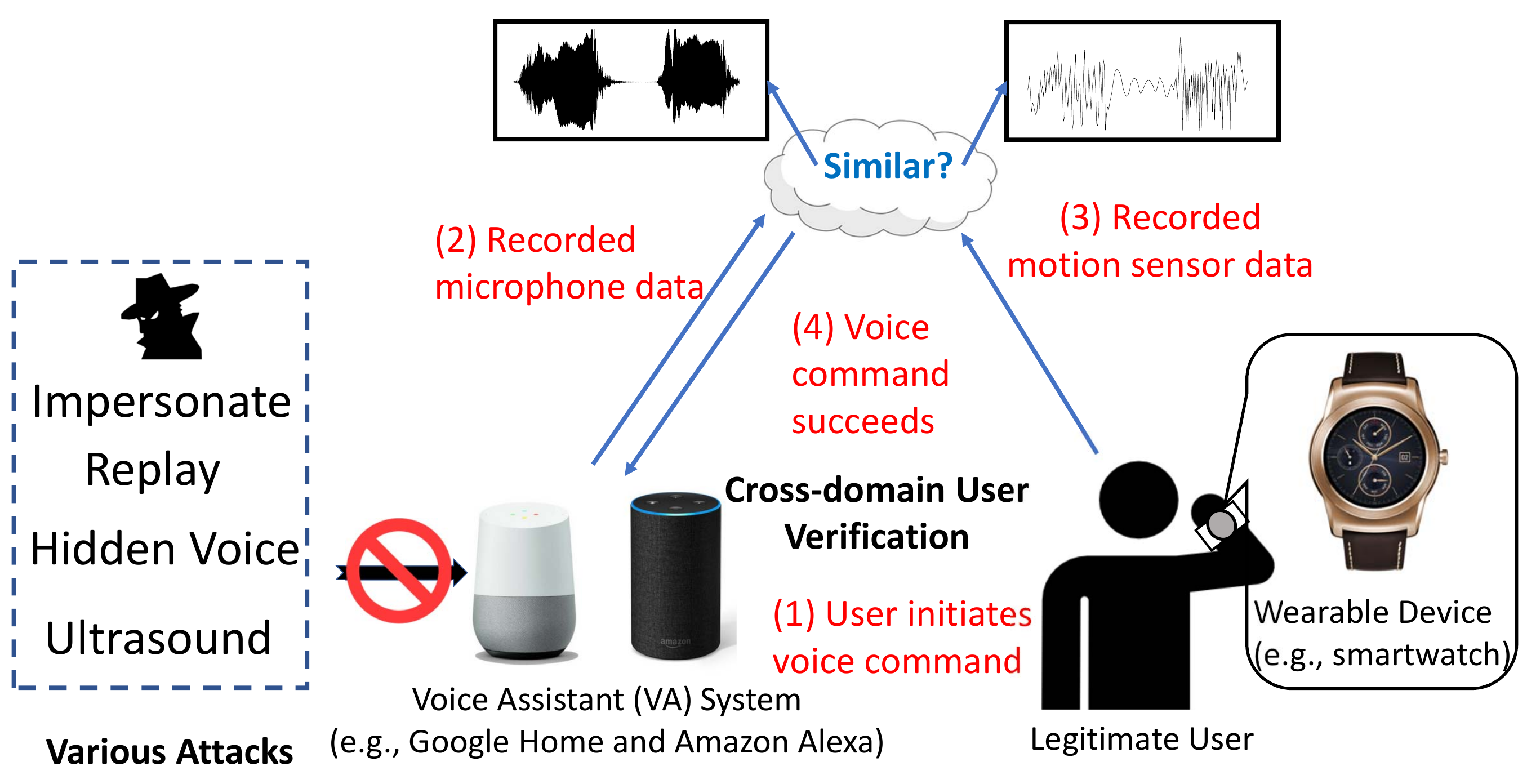}
 	\vspace{-4mm}
 	\caption{Proposed architecture of WearID.}
 	\label{fig:highlevel}
 	\vspace{-5mm}
 \end{figure}

	While the VA systems bring immense flexibility and convenience to users, the highly sensitive information collected by these systems could attract an adversary's interests and put the user's privacy under high risks.
	For instance, when the VA system is in an open area (e.g., in the office), the adversary can easily learn the user's schedule such as when to pick up his/her kid from daycare by asking the VA system ``What is my schedule to pick up my son''. Similarly, the user's private travel schedule such as when to attend a conference can also be easily revealed by requesting ``Remind me which day to attend the Machine Learning conference''.
	Therefore, both of the user's family and personal sensitive information can be obtained by simply asking one question.
	Furthermore, the adversary can even request the voice assistant to execute commands that are against the user's will.
	For example, the adversary can place online orders through the user's associated account without knowing the credit card information by telling the VA system ``Order a MacBook from Prime Now'', and then he can wait at the user's address to pick up the delivery.
	When the adversary can access the VA system at home remotely (e.g., through a hacked Smart TV), he can even use the voice command ``Unlock the exterior door'' to unlock the door's smart locking system and gaining entry into the house.

	Existing VA systems have deployed the voice biometric technology (at least to authenticate sensitive user commands), however, this approach identifies users based on their unique acoustic features solely in the \textit{audio domain} (i.e., extracting information from the data captured by microphones).
	The acoustic features are known to be
	vulnerable to \textit{impersonation attacks} and
	\textit{replay attacks}, where the adversary can fool the systems by imitating the legitimate user's voice~\cite{togneri2011overview} or via a simple record-and-replay of the user's previous voice commands~\cite{lindberg1999vulnerability}. Moreover, recent studies show that hidden audible voice attacks and even ultrasound attacks could access VA systems surreptitiously even when the legitimate user is present near the VA system~\cite{carlini2016hidden,zhang2017dolphinattack}.


To address the above vulnerabilities underlying VA systems, we propose a VA user authentication system, \textit{WearID}, which leverages the already pervasive wearable devices as an additional factor and performs cross-domain authentication on the user's voice commands.
WearID utilizes the low-cost motion sensors embedded in the user's wearable device to capture the unique voice characteristics in the vibration domain, which is compared to the same characteristics captured by the microphones in current VA systems and protect the VA systems from various audio attacks.
The flow of WearID is illustrated in Figure~\ref{fig:highlevel}.
Our system simply uses the regular \textit{wake word} (e.g., ``OK Google'') to trigger the authentication process. The command sound is then captured by the microphone of the VA system in the audio domain and the accelerometer of the wearable device in the vibration domain, respectively.

We develop a new algorithm to perform cross-domain comparison in audio and vibration domains. The software processing component is deployed in the VA system's cloud service to process the sensor data for user authentication.
If the similarity is high, the system accepts the voice command as from the legitimate user. Otherwise, the system rejects the voice command and sends a warning message to alert the user.
Our solution does not require special hardware or modifications to VA systems, therefore it can be easily integrated into existing VA systems and wearable devices. Compared to virtual buttons~\cite{virtualbutton2018}, our system is low-user-effort, which avoids tedious system training and cumbersome user operations on the mobile device in practice.

	Recent studies show the initial success of using motion sensors on the smartphone to capture the speaker's voice.
	For instance, Gyrophone~\cite{michalevsky2014gyrophone} presents that gyroscope can capture the acoustic signals from an external loudspeaker and reveal the speaker information (e.g., gender and identity).
	Accelword~\cite{zhang2015accelword} uses the smartphone's accelerometer to detect human voice for \textit{wake word} recognition.
	Speechless~\cite{anandspeechless} identifies the condition of using the smartphone motion sensor to capture sound: the shared hard surface between the external speaker and the smartphone.
	However, implementing WearID in practical scenarios using motion sensors in wearable devices to enhance the security of VA system is a challenging task.
	\textit{First}, the vibration domain information provided by the motion sensor and its unique acoustic characteristics remain unclear.
	\textit{Second}, the high-sampling-rate microphone data (e.g., 	8kHz and 44.1kHz) and the low-sampling-rate motion sensor data (e.g., 200Hz)
	are not directly comparable, the relationship between two distinct sensing
	modalities must be determined for a reasonable comparison. \textit{Third}, the
	synchronization of the two data sets from totally different hardware is
	difficult. \textit{Fourth}, the proposed system should defend against various audible impersonation and replay attacks~\cite{togneri2011overview, lindberg1999vulnerability} and inaudible attacks~\cite{carlini2016hidden,zhang2017dolphinattack}.

Toward this end, we explore the feasibility of leveraging the wearable's motion sensor to harness the aerial voice vibrations corresponding to live human speech. 
To ensure reliable cross-domain comparison, 
WearID develops a spectrogram-based method to convert the spectrogram of microphone data into that of lower-sampling-rate aliased signals, making it comparable to the spectrogram from motion sensor readings.
We extensively study the unique response distance and characteristics of the motion sensors in wearable devices and identify the complex relationship between the two sensing modalities to facilitate the data comparison. Our system is designed to maximize the usage of motion sensors' response in the frequency domain and focus on the acoustic signals with the frequencies and amplitudes that are perceivable to motion sensors during the microphone spectrogram conversion.
WearID leverages the VA system's wake word to trigger the verification process and start data collection on the VA and wearable devices simultaneously.
To trigger the data collection on the wearable device, WearID utilizes two alternative approaches based on WiFi communication or accelerometer-based wake-word detection and coarsely synchronize the two different sensing modalities. We develop a shift 2D-correlation method, which shifts the spectrogram of the two sensing modalities' readings within a short time window to reduce the residual synchronization errors and obtains the maximum 2D correlation to describe the cross-domain similarity. In addition, WearID calibrates the data to remove the vibration noises (e.g., hand motions) and identify precise command sound segment during data preprocessing.
This proposed system eventually reveals the unique relationship between two types of signals, which contains rich information embedded in the spectrograms in two domains those are hard to be forged by adversaries and make WearID resilient to various attacks including audible impersonation and replay attacks and inaudible attacks.

\smallskip
\noindent \textbf{Our Contributions:} 
	\begin{itemize}[leftmargin=*]
		\item We find that human voices can be captured over the air by the motion sensors embedded in wearable devices. This could serve as an additional domain (i.e., vibration domain) to the original audio domain to verify the user and secure the VA system.
		\item We propose a unique cross-domain user verification system, WearId, which requires low-user-effort and no modifications to the existing VA systems; it can be easily integrated with existing VA systems and wearable devices.
		\item The motion sensor's response to voice signals in short range offers advantages to effectively prevent the impersonation and replay sounds from accessing the wearable. We derive the unique spectrogram relationship between two sensing modalities (i.e., microphones and motion sensors) to provide enhanced user verification using wearable devices.
		\item We conduct extensive experiments and user studies with different models of smartwatches and participants, which result in $600$ human voice segments. The results show that WearID can authenticate user's voice commands with $99.8\%$ accuracy in the normal situation and detect $97\%$ of various impersonation and replay attacks with a low false negative rate of $2\%$. When under the hidden voice and ultrasound attacks~\cite{zhang2017dolphinattack}, WearID achieves close to $100\%$ accuracy of verifying the users.
	\end{itemize}
%

\textbf{Why Use a Wearable-Assisted Approach?}
The convenience of the wearable device has greatly contributed to its wide deployment in these years. The number of wearable devices worldwide reaches $593$ million in 2018~\cite{wearable2018}. Due to its wearable/coexisting nature with their owners, the wearable devices have been considered an effective security token in various payment systems~\cite{murray2017top}.
	In this work, we find that the motion sensors widely used in wearable devices can capture people's speech through the vibration in the air
	at a short distance (e.g., $25$cm to the mouth), which is favorable for rejecting the spoofing attacks because it is hard to be so close to the user's wearable.
	We also find that the vibration-domain information of the human voice contains unique characteristics to distinguish different users.

	\vspace{-2mm}
\section{Related Work}
\label{sec:related}







\textbf{Audio-domain Voice Authentication and Security Issues.}
The traditional user authentication methods designed for voice access systems mainly extract and distinguish each individual's voice features in the audio domain to identify users~\cite{wechatvoiceprint, Siri, hebert2008text, variani2014deep, campbell1997speaker, reynolds1995robust}.
For example, Mel-Frequency Cepstral Coefficients (MFCCs)~\cite{murty2006combining} and Spectral Subband Centroids (SSCs)~\cite{kinnunen2007speaker} describe a voice's timbre and vocal-tract resonances and are widely used as unique voice features to distinguish users. The modulation frequency~\cite{atlas2003joint} capturing formant and energy transition details of a voice sound contains speaker-specific information for user identification.
Although these voice authentication approaches seem to be convenient, they only rely on the audio-domain features which are vulnerable to acoustic-based attacks.
For example, an adversary can spoof the legitimate user to pass the voice authentication system by recording and replaying a user's voice sound~\cite{lindberg1999vulnerability}.
In addition, the adversary can study the user's daily speeches and impersonate or synthesis the user's sound to pass the voice access system~\cite{togneri2011overview, lindberg1999vulnerability, de2010evaluation,de2010evaluation}.
\textbf{WearID Versus Other Authentication Methods.}
To defend against the replay and impersonation attacks, researchers show that advanced speaker models, Gaussian Mixture Model and i-vector models~\cite{amin2013detecting,hautamaki2013vectors}, and the speech features, relative phase shift and modulation features~\cite{de2012evaluation,wu2013synthetic} could be used to defend the voice authentication systems.
However, these solutions solely use the features from the audio domain, which are still vulnerable to audio-based attacks if the attackers have the knowledge of the characteristics of these features.
Recently, more researchers propose to determine the liveness of the sound source
by exploiting the physical features of human speeches other than the voice features~\cite{chen2017you,zhang2016voicelive, zhang2017hearing}.
Specifically, Chen~\textit{et al.}~\cite{chen2017you} examine the unique magnetic field patterns generated by electro-acoustic transducers to detect whether the voice sound is generated by a loudspeaker or not.
VoiceLive~\cite{zhang2016voicelive} and VoiceGesture\cite{zhang2017hearing} derive the time-difference-of-arrival (TDoA) and the Doppler shifts from the received sound and detect the dynamic acoustic characteristics that only occur in human sound to detect liveness.
However, these approaches are focusing on smartphone and require the user to place the smartphone or microphone close to the mouth or loudspeaker. Thus they are not applicable to the VA systems (e.g., Google Home and Amazon Alexa) that allow users to give voice commands from distance. Feng~\textit{et al.}~\cite{feng2017continuous} develop a user verification system that can defend against impersonation, replay and synthesis attacks when using the VA systems. The developed system captures the user's facial vibrations via an accelerometer embedded in a pair of glasses. The vibrations are then compared with the voice sound recorded by the VA system to verify whether the voice command is given by the user wearing the glasses. However, this approach requires the user to wear a dedicated device with a high sampling-rate accelerometer and needs to modify the VA device hardware.

\textbf{Vibration-domain Voice Recognition.}
Recent studies show that the MEMS motion sensors (e.g., accelerometer and gyroscope) are able to capture acoustic sounds~\cite{michalevsky2014gyrophone, zhang2015accelword, crager2017information}. Gyrophone~\cite{michalevsky2014gyrophone} utilizes the gyroscope in a smartphone to recognize the speaker's information (e.g., gender and speaker identity) from the speech played by a loudspeaker. Accelword~\cite{zhang2015accelword} leverages the accelerometer in a smartphone to recognize the user's wake word sound, which reduces the energy consumption of the personal VA system in the smartphone (e.g., Siri). Speechless~\cite{anandspeechless} further analyzes the speech privacy leakage including the speech content from the smartphone motion sensors under various attacking scenarios.
These works only focus on motion sensor measurements and do not reveal the relationship between the sensor readings and real voice recorded by microphones. Moreover, these works only prove that human voice could impact motion sensors in smartphones or dedicated wireless sensors, but the impact to the motion sensors in wearable devices attached to human bodies is unknown.

In this work, we build the first cross-domain authentication system, WearID, which can verify the legitimacy of the voice commands received by a VA system through the unique relationship of the human voice in the vibration and audio domains. Different from the existing works, we show that the motion sensors embedded in commodity wearable devices can capture the aliasing components of the human voice, which have distinct characteristics that can facilitate voice authentication. Our WearID leverages the fact that human voice result in the mechanical waves in the air, which can be captured by the sensing modalities in both audio and vibration domains.
Our solution requires low user efforts and minimum modification to existing VA systems. The system is training-free and does not store users' biometrics.

	\vspace{-1mm}
\section{Vulnerabilities of Voice Assistance Systems}
\label{sec:preliminary}


\vspace{-2mm}
\subsection{Potential Security Breaches in VA Systems}
\sloppy
While the VA system brings great convenience, flexibility, and multi-functions to users, the open nature of the voice access to the VA systems (i.e., anyone can access the VA systems via voice sounds) could cause serious security breaches.
We study current VA systems' types, related at-risk information/operations, and their limited defense methods.
\textbf{Two Types of VA Systems.}
Based on whether the VA system is shared among a group of users or not, we divide the current commodity VA systems into two types:
\textit{Personal VA Systems} are designed only for personal use. They are usually integrated into users' mobile devices, such as smartphones (e.g., Google Now and Siri).
Differently, \textit{Family/community Shared VA Systems} are designed to be used in the home or office environments. They are usually built into a stand-alone device and shared by multiple users. The typical commodity products of this type are Amazon Alexa and Google Home.
The differences between these two types of VA systems are that the personal VA system usually takes voice commands using the microphone of the user's mobile device, which is in the proximity to the user, while the family/community shared VA system is usually designed to pick up users' voice commands from a distance in a house or office. The family/community shared VA systems are considered to be of higher risk because adversaries could easily access the VA systems without being noticed. Thus, this type of VA system is the primary focus of this work.

\textbf{At-risk Information/Operations in VA Systems.}
VA systems are usually linked to the users' personal information and even their family/community information. When an adversary has access to the VA system, he can easily get the user's private information. For example, the adversary can get the user's shopping information by asking ``What is in my shopping list?''. The adversary can also get the personal email content by saying ``Read me my email''. Furthermore, the adversary can get the user's family schedule via the voice command ''List all events for January 1st''.
Moreover, recent VA systems are usually deployed as a hub connecting various smart appliances at home or in the office. In that case, the adversary can use voice commands to control the smart appliances without permission. For instance, an adversary can put the user in danger by saying ``Unlock the exterior door''.
Along with this direction, we investigate the privacy-sensitive voice commands, related to issues of private information leakage and unauthorized operations.
\textbf{Limited Defense Methods in VA Systems.}
Most of the off-the-shelf VA systems require a pre-defined voice command (known as \textit{wake word}) to wake up the system, such as ``Alexa'' and ``OK Google''.
These wake words could be used to verify the identity of the speaker by comparing with the pre-recorded sounds in the user profile, which is built when the user enrolls the system.
However, such audio-based speaker verification in the current VA systems is not trustworthy, because the acoustic features they depend on can be easily spoofed. To warn the VA system users about this issue, Google Home particularly notes that \textit{``A similar voice might be able to access this info, too''}~\cite{google2018set}.
Furthermore, current VA systems can only verify users based on the wake words, leaving the voice commands unprotected from the attacks. Thus an adversary only needs to focus on attacking the wake words, which makes the attack much easier. In addition, we find that the VA system stays in the listening mode for a long time (e.g., $30$ seconds for Google Home) to capture voice commands after being woken up. During this time period,
the VA system is defenseless to any adversary.
Moreover, research shows that the audio-based VA systems are vulnerable to various audio attacks, including imitation and replay attacks.


\vspace{-2mm}
\subsection{Attack Model}

We consider an adversary who is interested in obtaining the user's private information or exerting an unpermitted operation from the stand-alone family/community shared VA device, which involves multiple users and is exposed to much more security issues compared to the personal VA device. We assume the adversary can not physically break the VA device, take control of the VA cloud service or get the possession of the user's wearable device. We also assume that the VA user always wears a wearable when using the VA system, which is normal for most wearable users. We summarize the potential attacks in two major categories. \textit{Attack on User's Absence} is one type of attacks that need to be launched when the user is absent from the VA device. Otherwise, the user could notice the attacking sound and stop the attack. \textit{Co-location Attack} is the other type of attacks that can be launched surreptitiously even when the user is present to the VA device without causing notice.

\textbf{Attack on User's Absence.}
When the user is away from the VA device, an adversary could get close to the VA device and launch the following attacks without being detected:

\begin{itemize}[leftmargin=*]
	
	\item Random Attack. An adversary who does not know the user's voice characteristics can try to fool the VA system by using his own voice. Because the adversary only needs to attack the single wake word, such attack still has high success rates. In addition, this attacking scenario includes occasional access. For example, a family/community member having similar voice characteristics may access another user's private information accidentally.
	
	\item Impersonation Attack. An experienced adversary who knows the user's voice characteristics can attack the VA system by imitating the user's voice. The adversary can also synthesize the user's voice by using an audio editing software and playback the synthesized sound via a loudspeaker to launch the attack.
	
	\item Replay Attack. An adversary who has the opportunity to observe the user's voice command sounds can use a microphone to record the voice sounds and playback the recorded voice commands via a loudspeaker to fool the VA system.
	
\end{itemize}

\textbf{Co-location Attack. }
When the user is in proximity to the VA system, an adversary 
could still launch the following attacks without being detected:

\begin{itemize}[leftmargin=*]
	\item Hidden voice attack. An adversary may embed the recorded user's voice commands into the background of music or video streams or directly generate command sounds according to the knowledge of the underlying VA system~\cite{carlini2016hidden}. The generated command sound can be recognized by VA systems but not perceptible to human. Moreover, an adversary can control the volume or mute the VA device via hidden commands to avoid being noticed from the audible reply.

	\item Ultrasound Attack. An adversary may use a microphone to record the user's voice commands, modulate the recorded voice commands onto the ultrasound frequency band (i.e., $\geq20KHz$), and use the modulated sound to fool the VA system. Although human ears can not hear the modulated voice commands, they can still be recognized by existing VA systems~\cite{zhang2017dolphinattack} due to the non-linearity of the microphone.
	
\end{itemize}

\section{User Verification Design}
\label{sec:system}

\subsection{System Purpose and Challenges}

Our system checks the proximity between the user and his/her voice command sound to verify whether the voice command comes from the legitimate user, who is also present to the VA system. The basic idea is to compare the command sound across the audio domain (i.e., via the VA device's microphone) and the vibration domain (i.e., via the motion sensor of the user's wearable). If the command sound matches across the two domains, the voice command is verified to come from the legitimate user (i.e., owner of the wearable device).
Existing VA systems only focus on verifying wake words and neglect the protection of more sensitive voice commands, which opens more opportunities for adversaries.
In comparison, our system requires confirming whether the voice commands come from the right user.
There are many challenges to design such a system.
For example, it is challenging to match the command sound from the sensors working in two different domains, which have a considerable gap in sampling rates (e.g., 8000Hz versus 200Hz).
In addition, it is unknown whether the motion sensors on wearables provide sufficient information to characterize human voice sounds, in light of their low fidelity and not dedicated purpose for recording sound.
Furthermore, how to trigger synchronize the authentication process on both the VA device and the wearable device need to be explored.
As last, the proposed authentication system should defend against various attacks when the user is present or absent to the VA device. 

\begin{figure}[t!]
	\centering
	\includegraphics[width=2.9in]{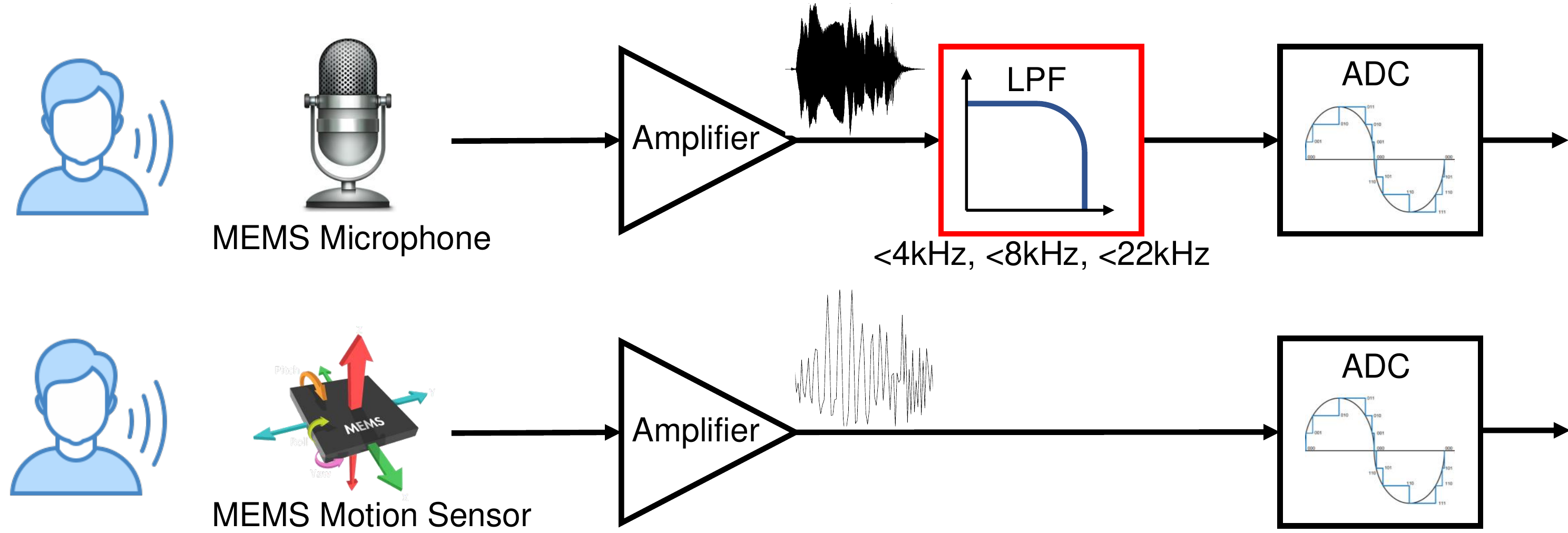}
	\vspace{-4mm}
	\caption{Hardware flow of microphone and motion sensor.}
	\label{fig:preliminary}
	\vspace{-7mm}
\end{figure}

\vspace{-1mm}
\subsection{System Flow}
\begin{figure}[t]
	\centering
	\includegraphics[width=3.0in]{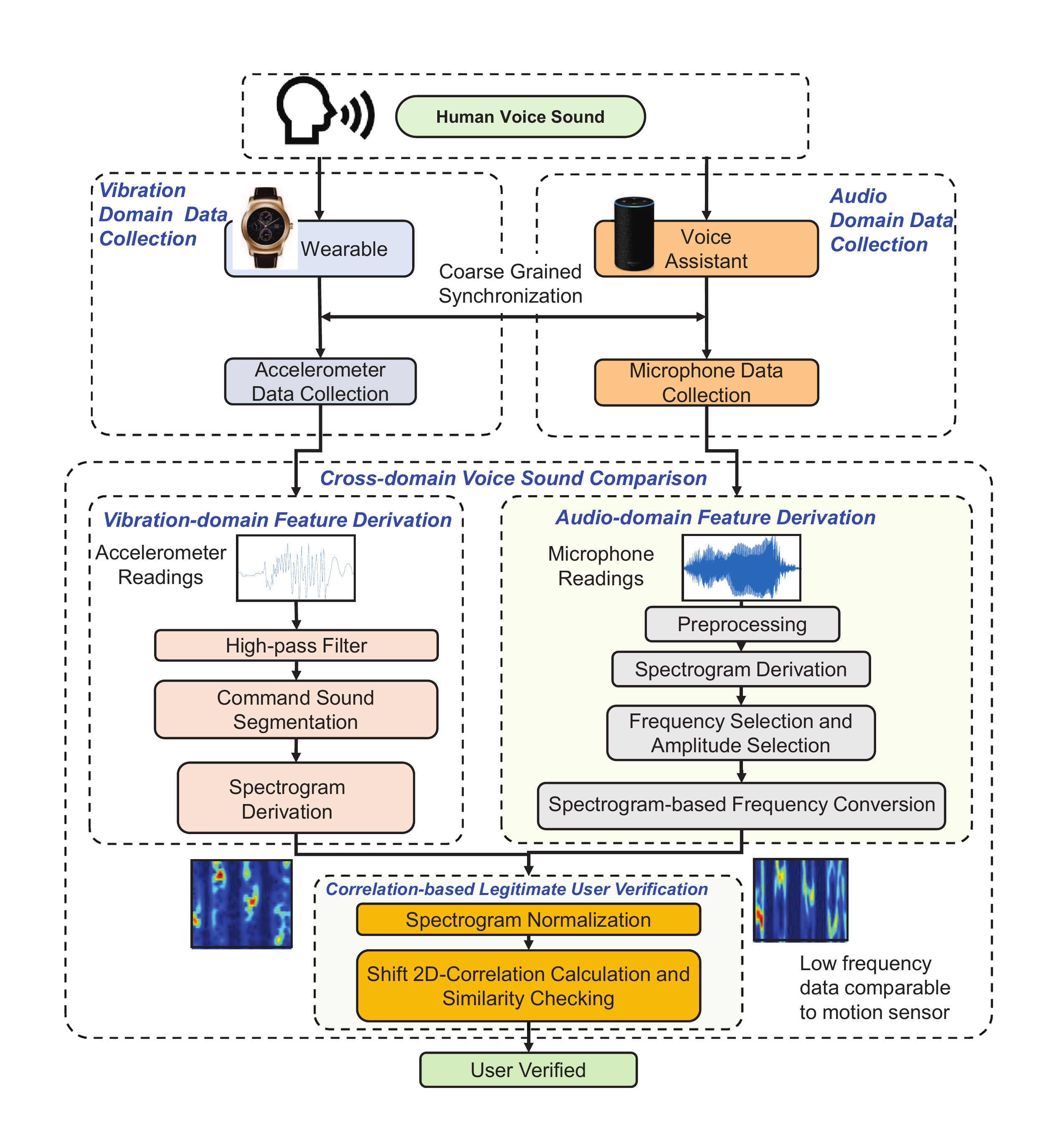}
	\vspace{-3mm}
	\caption{User verification overview.}
	\label{fig:system}
	\vspace{-7mm}
\end{figure}

Toward this end, we develop a unique VA system, WearID, which verifies the authenticity of voice commands through the sensing data obtained from the vibration and audio domains.
Figure~\ref{fig:system} illustrates the flow of WearID.
Our system requires the \textit{Audio Domain Data Collection} and the \textit{Vibration Domain Data Collection} to collect the microphone from VA device and accelerometer data from the wearable, which describe the user's voice interaction with the VA system in two different domains.
\textit{Coarse-grained Synchronization} aims at triggering the motion sensors to record the voice commands at the right time (i.e., after the wake word) and provides coarse synchronization between the two domain data.
The insight is that the wake word, which is mandatory for initiating the VA device, could be utilized to trigger the data collection on both devices at the same time. We propose two alternative approaches to achieve the coarse-grained synchronization. The \textit{WiFi Communication-based approach} only requires the VA device to detect the wake word and trigger the wearable to start data collection through the WiFi communication when both devices are in the same WiFi network~\cite{androidwear2018}.
We note that emerging wearables are in a trend of having standalone WiFi modules that can connect to WiFi networks directly. In the case of wearables not having WiFi modules, they still can connect to WiFi networks through their paired smartphones.
The alternative approach \textit{Parallel Wake-word Detection-based approach} uses the motion sensor in the wearable device to detect the wake word independently based on voice recognition in vibration domain to initiate the sensor data collection. To achieve this, WearID reuses the motion sensor data from the ongoing fitness tracking App, which continuously counts the user's walking steps. 
After the synchronization, both the wearable and VA device start to continuously collect the accelerometer and microphone data respectively to capture the voice command. The data will be uploaded to a cloud server that is running our processing algorithm to further derive the cross-domain similarity for user authentication.
WearID has three core components: The \textit{Vibration Domain Feature Derivation} and \textit{Audio Domain Feature Derivation} derive reliable time-frequency features from the data collected by the wearable's motion sensor and the VA device's microphone, respectively. The derived features are converted to comparable spectrograms based on a complicated unique relationship between the audio and vibration domains. The \textit{Correlation-based Legitimate User Verification} calculates the similarity between the spectrograms from two domains for user verification.

In particular, the \textit{Vibration Domain Feature Derivation} removes the mechanical noise (e.g., due to hand movements) from the motion sensor data by using a high-pass filter and extracts the command sound segment from the motion sensor readings by examining the moving variances. The two-dimension time-frequency description of signal, spectrogram, is then derived from the identified motion sensor segment. Similarly, the \textit{Audio Domain Feature Derivation} pre-process the microphone data to remove the acoustic noise and identify the command sound segment, which is utilized to derive the spectrogram. The next is to convert the microphone spectrogram to the low-frequency form comparable to the motion sensor and maximize the intersection of the two distinct sensing modalities' acoustic responses. The \textit{Frequency Selection and Amplitude Selection } select the frequency/amplitude of the microphone spectrogram for conversion according to the identified unique acoustic characteristics of motion sensor. Moreover, the \textit{Spectrogram-based Frequency Conversion} converts the microphone spectrogram into low-frequency spectrogram, which describe the similar responses on the motion sensor. The \textit{Correlation-based Legitimate User Verification} first performs the \textit{Spectrogram Normalization} to normalize the time lengths and magnitudes of the spectrogram in two domains. \textit{Shift 2D Correlation-based Similarity Calculate} computes the 2D-correlation between the spectrograms to check the similarity and shift one spectrogram over time during calculation to address the synchronization errors. The resulted maximum 2D-correlation coefficient is compared with a threshold to determine whether the command sound received by the VA device is from the legitimate user (i.e., wearable owner).

\begin{figure}[t!]
	\begin{center}
		\begin{tabular}{cc}
			\includegraphics[width=1.3in]{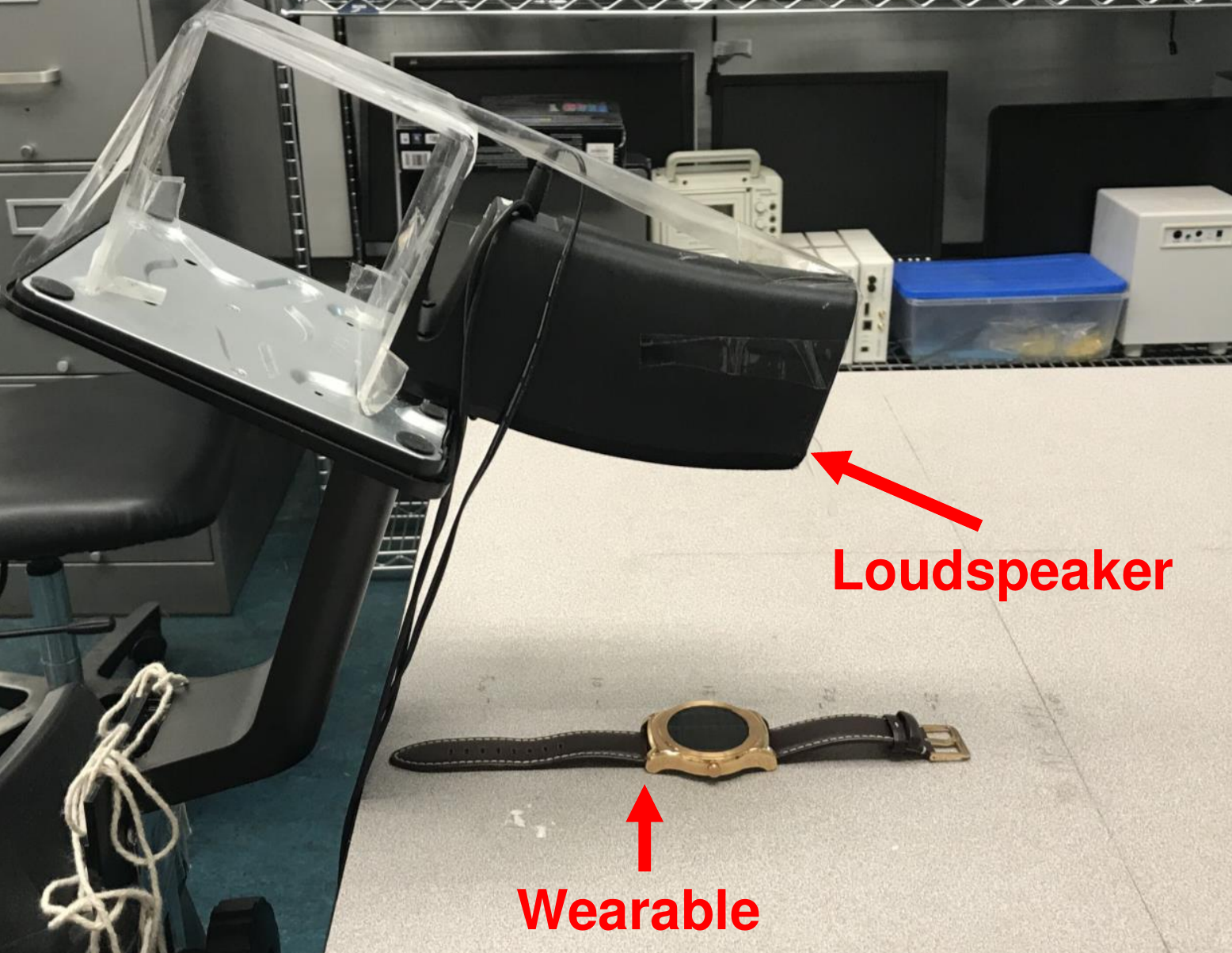}
			&
			\includegraphics[width=1.5in]{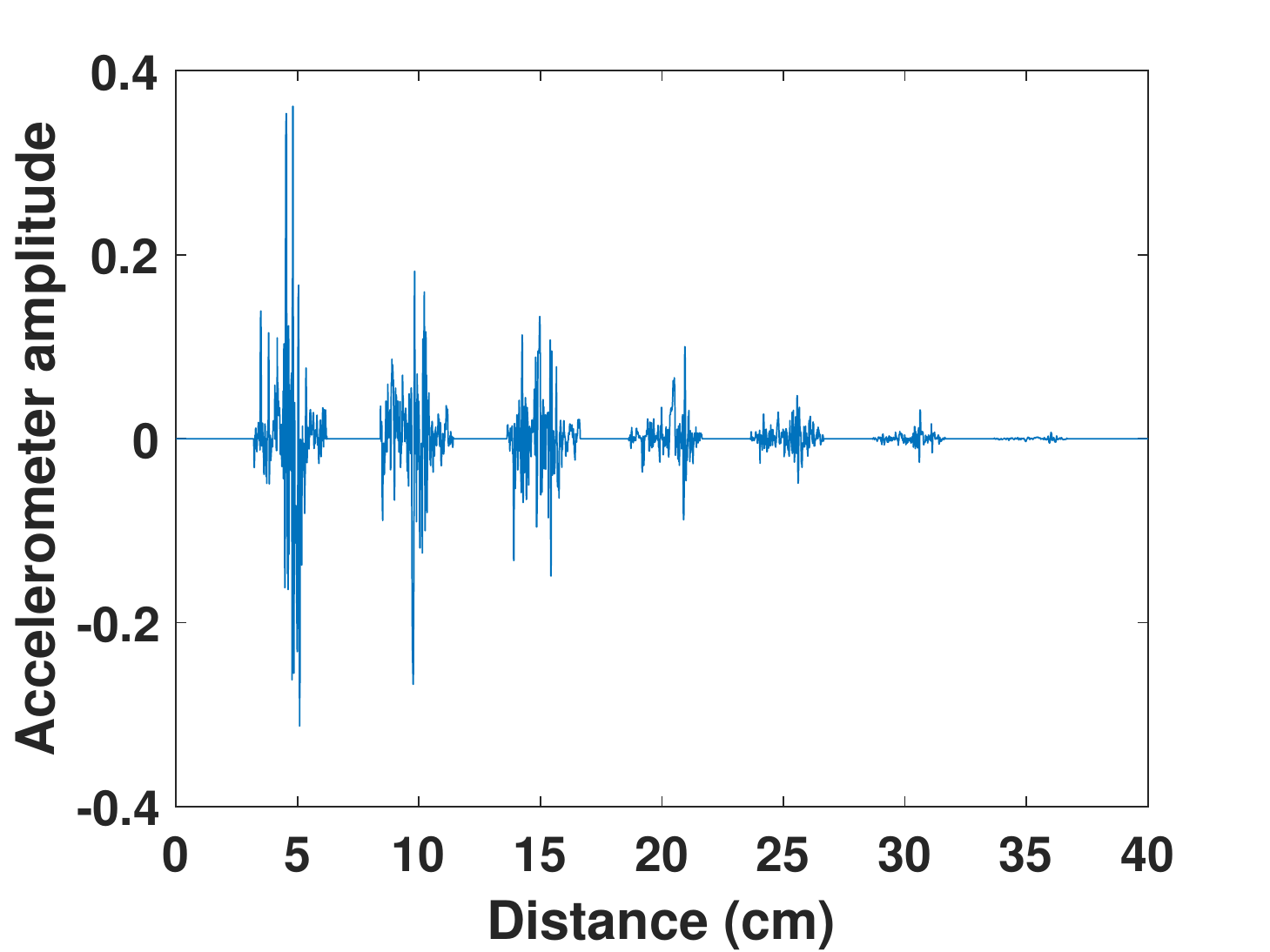}
			\\{\scriptsize(a) Experiment setup} & {\scriptsize(b) Response distance test}
		\end{tabular}
	\end{center}
	\vspace{-4mm}
	\caption{Experiment setup for preliminary study and the accelerometer response distance test.}
	\vspace{-5mm}
	\label{fig:preliminary_setup}
\end{figure}
\vspace{-2mm}
\subsection{Why Wearable? Why Motion Sensor? Why not a Second Microphone? }

We consider the wearable device as a trusted device because of its unique attribute, usually worn on the user body and rarely left unattended. In addition, many recent wearable-based continuous authentication schemes further guarantee its usage as a unique ID. Moreover, the pervasive deployment of the wearable device further facilitates its usage as an ID. The wearable devices are embedded with many sensors, such as the microphone and the motion sensor. Why does WearID choose motion sensors to provide enhanced security to the VA system instead of a microphone on the wearable? The reasons are three folds. First, only a small number of the high-end wearable devices contain a microphone. In comparison, most of the wearable devices are embedded with a motion sensor due to its design purpose for various applications such as fitness tracking, gesture control and activity recognition. Second, the motion sensor has a short response distance to sound, which prevent most acoustic accesses that are from the sound source other than the wearable owner. Three, motion sensors provide a distinct sensing modality and the user verification across two domains provides much higher security than adding an additional microphone. If an adversary can issue impersonated commands to the voice assistant's microphone, it can also do the same to the second microphone, if used. For example, the two-microphone based approach is vulnerable to ultrasound and hidden command attacks~\cite{blue20182ma}. Fourth, the low-frequency motion sensors have low power consumption and do not require much programming or processing on the wearable.

\vspace{-1mm}
\section{Feasibility of User Verification in Vibration Domain}
\label{sec:preliminarystudy}
\vspace{-1mm}
\subsection{Relationships and Differences between Microphone and Accelerometer}
\label{subsec:hardward}
Both microphone and accelerometer are Micro Electro Mechanical System (MEMS) sensors.
However, the hardware designs of the two sensors are greatly different. We compare the processing components reside in microphone and accelerometer as the following.

\textbf{MEMS Microphone.}
The MEMS microphones are widely used in various sound recording devices, including voice assistants (e.g., Google Home and Amazon Echo). The MEMS microphone consists of a membrane and a complementary perforated black-plate~\cite{wang2016windcompass}. In the presence of human voice, the sound passes through the holes in the black-plate and hits the membrane, where the sound waves are captured according to the capacity changes and are converted to analog signals~\cite{fuller2005microelectromechanical}.
Figure~\ref{fig:preliminary} shows the hardware components in a microphone.
We can find that the analog signals are amplified and fed to a Low Pass Filter (LPF), where the cutoff frequency is set to 4KHz, 8KHz or 22KHz corresponding to the device's sampling rate 8KHz, 16KHz and 44KHz. The Analog-to-Digital Converter (ADC) then digitizes the analog signals.
Particularly, the sample rate of ADC determines the maximum frequency of the recorded sounds, though the analog signal can capture sound with higher frequencies (i.e., over $22kHz$).

\textbf{MEMS Accelerometer.}
The accelerometers used in most mobile devices and wearable devices are MEMS also sensors.
In the accelerometer, the sound could be measured as the subtle movements of the inertial mass caused by the changing sound wave pressures.
The set of the hardware components of the accelerometer is different from that of the microphone. As shown in Figure~\ref{fig:preliminary}, the accelerometer does not contain a LPF between the amplifier and the ADC. Thus it would experience high signal aliasing if the signal before the ADC is greater than the sampling rate (e.g., the maximum sampling rate $200Hz$ allowed in Android system).
Such characteristics, however, could be exploited to capture sounds with frequencies higher than the accelerometer's sampling rate.

\vspace{-1mm}
\subsection{Response Frequency and Distance of the Built-in Accelerometer}


Though the motion sensors have been shown to be resonant to sounds, when they are embedded in the wearable devices, their responses to sound are also affected by the device cases and other electric components. To examine the capability of built-in accelerometer on picking up the sound, we conduct a set of analysis for its response distance and frequency.

\textbf{Short Response Distance.} We first validate the capability of built-in accelerometer on picking up voice under various distances. A preliminary experiment is conducted leveraging a loudspeaker, which plays a recorded voice command (i.e., "one") with the fixed volume. As shown in Figure~\ref{fig:preliminary_setup} (a), we place the loudspeaker on a stand, which does not share any solid surface with the wearable (i.e., LG Urbane W150). The voice command is played under distances from $5cm$ to $35cm$. In Figure~\ref{fig:preliminary_setup} (b), we can observe that the amplitude response of accelerometer decreases with the distance, and over the distance of $25cm$, the response to sound can be barely observed. Such short response distance of the accelerometer can help to reject voice commands that are not from the wearable owner.

\begin{figure}[t]
	\begin{center}
		\begin{tabular}{cc}
			\includegraphics[width=1.5in]{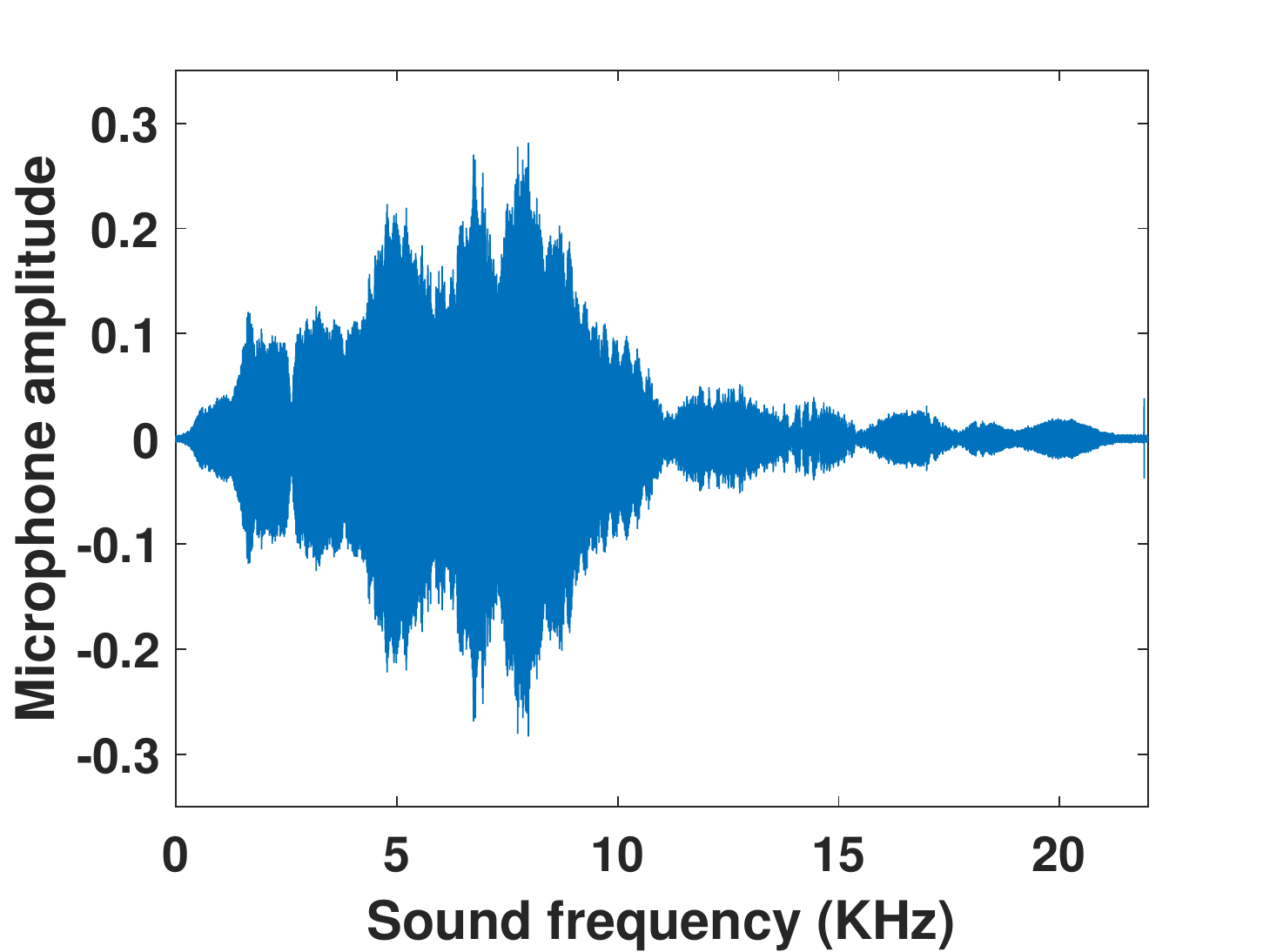}
			&
			\includegraphics[width=1.5in]{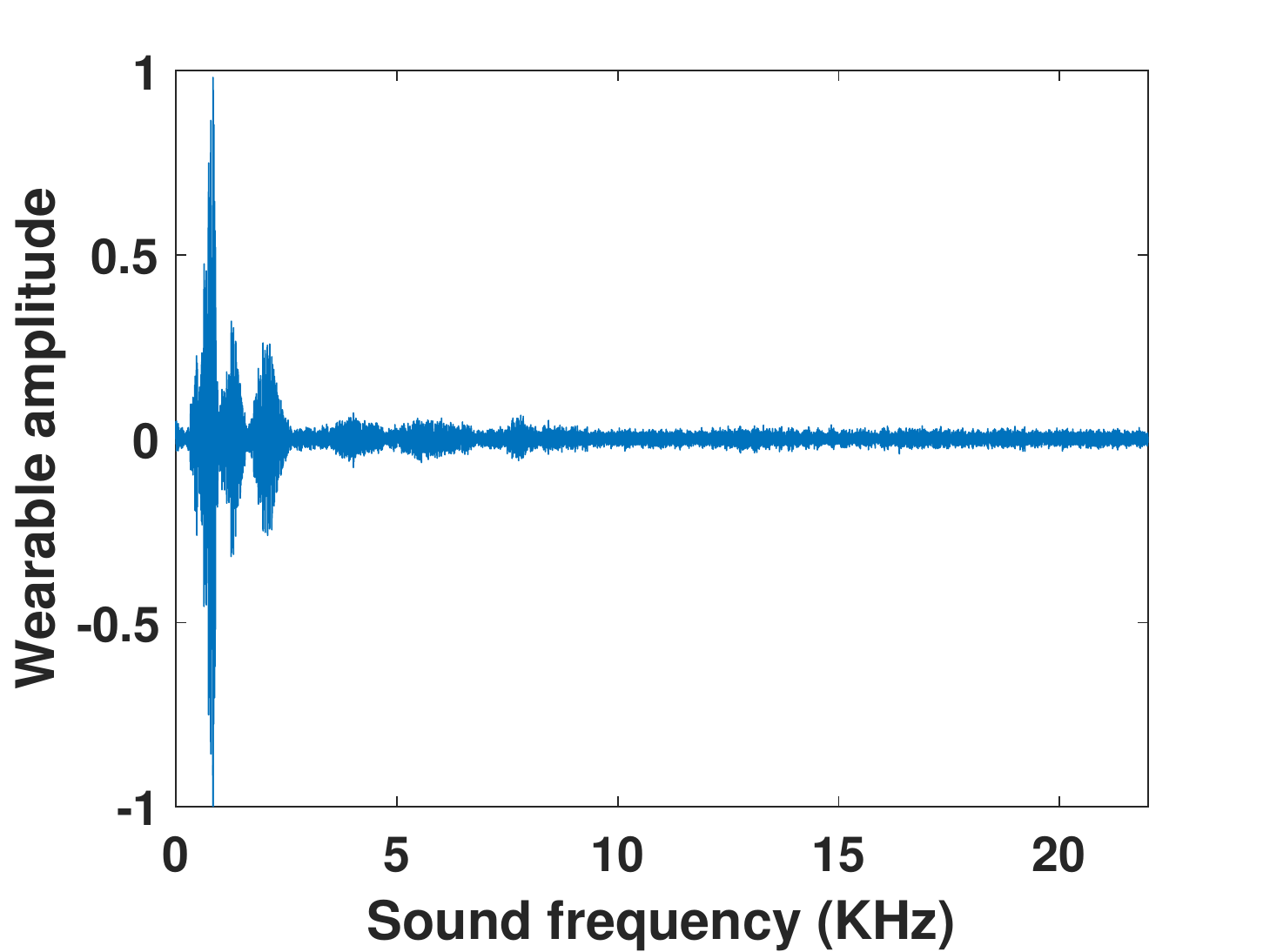}
			\\{\scriptsize(a) Amplitude of microphone} & {\scriptsize(b) Amplitude of wearable accelerometer}
		\end{tabular}
	\end{center}
	\vspace{-4mm}
	\caption{Responses of the microphone and the accelerometer to a chirp from 0Hz to 22kHz (in time-domain amplitude).}
	\vspace{-5mm}
	\label{fig:nonlinearity_chirp_0_22_time}
\end{figure}

\textbf{Frequency Response.} Next, we examine the built-in accelerometer's frequency response to the sound. We generate a frequency sweeping signal from $0Hz\sim{22kHz}$ and play the sound with a loudspeaker. An accelerometer (i.e., on Huawei Watch 2) and a microphone (i.e.,on Nexus 6) are placed on the desk, which does not share any solid surface with the loudspeaker as shown in Figure~\ref{fig:preliminary_setup} (a). We show the frequency responses of the two devices in Figure~\ref{fig:nonlinearity_chirp_0_22_time}.
We observe that the wearable's accelerometer shows a shorter range of frequency response to the sound, compared to the microphone.
The results are promising since the wearable device's accelerometer can capture sounds with frequencies that the major human voice resides in.
Additionally, the wearable's limited capability on sensing the high-frequency sound (i.e., over $4kHz$) show the potential to reject the audible hidden voice attack~\cite{carlini2016hidden} and the inaudible ultrasound attacks~\cite{zhang2017dolphinattack}.

\begin{figure}[t]
	\begin{center}
		\begin{tabular}{cc}
			\includegraphics[width=1.3in]{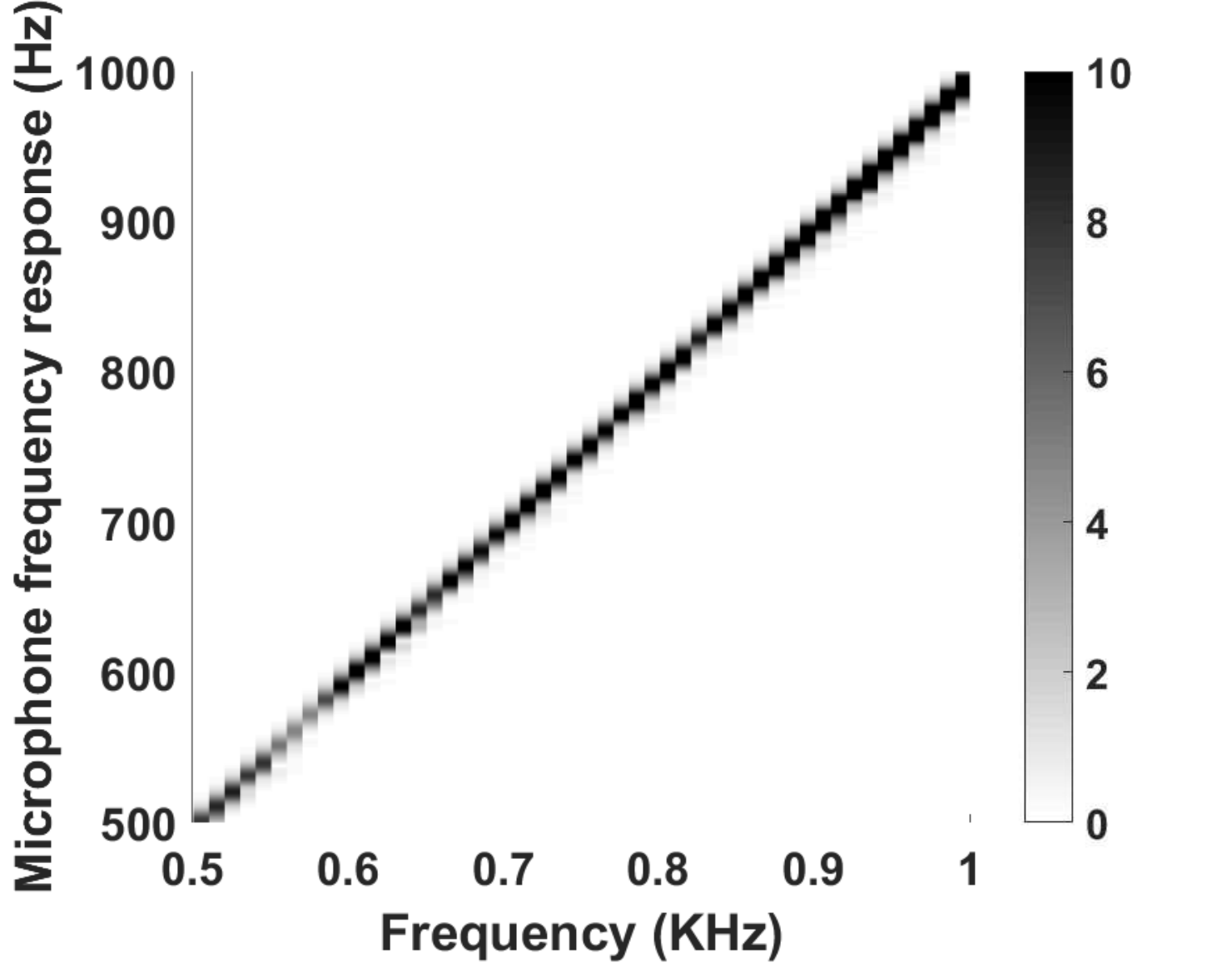}
			&
			\includegraphics[width=1.3in]{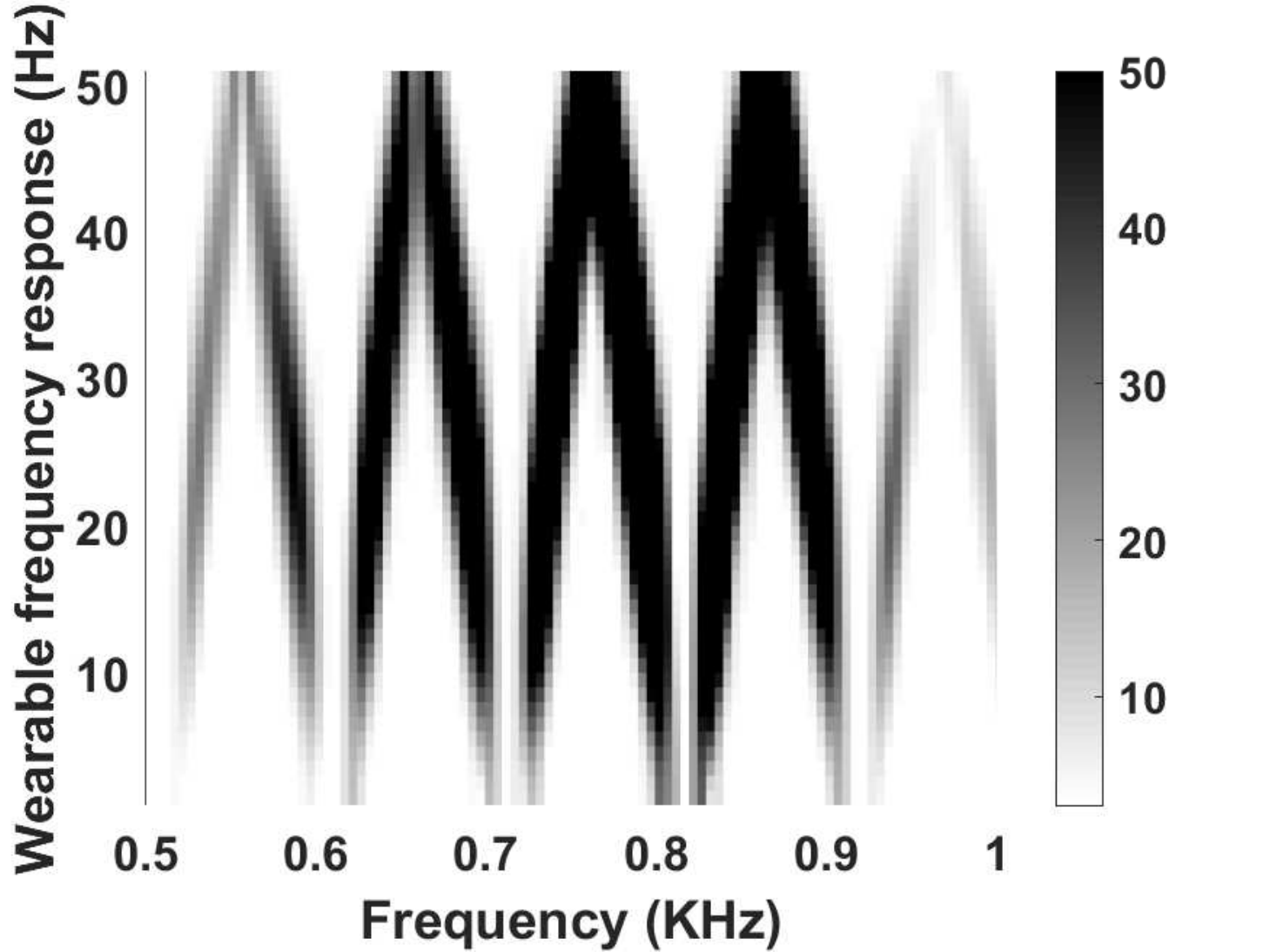}
			\\{\scriptsize(a) Spectrogram of microphone} & {\scriptsize(b) Spectrogram of wearable accelerometer}
		\end{tabular}
	\end{center}
	\vspace{-4mm}
	\caption{Frequency response of chirp $500Hz\sim1000Hz$ for microphone and accelerometer of wearable device (i.e., Huawei watch 2 sport).}
	\vspace{-7mm}
	\label{fig:nonlinearity_chirp}
\end{figure}

\vspace{-1mm}
\subsection{Interpreting Accelerometer Readings via Signal Aliasing}


The wearable device's motion sensor sampling rate is limited within $200Hz$ by the operating system. According to Nyquist theorem, the sampling rate of $200Hz$ allows the motion sensor to present the signal with up to $100Hz$ frequency. As introduced in Section~\ref{subsec:hardward}, the lack of a LPF before the ADC causes the motion sensor to exert signal aliasing when receiving sound signals. But due to this, the motion sensor can go beyond its sampling rate to hear the sound frequency higher than $100hz$. The signal aliasing is a phenomenon when the frequency components shift to new frequency points and overlapped with that frequency.
Particularly, we describe the aliasing signal as following:
\begin{equation}
\setlength{\abovedisplayskip}{3pt}
\setlength{\belowdisplayskip}{3pt}
\begin{aligned}
f_{alias}=|f-Nf_{s}|, N\in{Z},
\end{aligned}
\label{eq:nonlinearity}
\end{equation}
where $f_{alias}$, $f$ and $f_{s}$ denotes the aliasing frequency, original audio signal frequency and sampling rate. This equation reveals the relationship between the microphone data and the motion sensor data. In particular, for each audio frequency $f$, we can calculate the response frequency point $f_{alias}$ at the wearable's accelerometer and we can reconstruct the frequency responses of motion sensor as Figure~\ref{fig:nonlinearity_chirp}(b), if given a chirp microphone data as Figure~\ref{fig:nonlinearity_chirp}(a). However, Equation~\ref{eq:nonlinearity} shows that one frequency point at the motion sensor could be resulted from many audio frequencies. And the combination of many audio frequencies such as the human voice sound could make the aliased signal on the wearable very hard to interpret. Thus in this work, one of my major task is to explore the unique relationship between the microphone data and the accelerometer data and match the sound across the two domains for authentication.

\vspace{-1mm}
\subsection{Challenge of Cross-domain Command Sound Verification}

WearID matches the command sound across the audio domain and the vibration domain to verify the voice commands. However, due to the huge sampling rate gap between the two sensing modalities and their distinct characteristics when responding to sound, the cross-domain sound comparison is not trivial.
Specifically, the sampling rates of microphone data are normally above $8kHz$ while the rates for the wearable motion sensor is at most $200Hz$.
Re-sampling the microphone data and the motion sensor data to the same sampling rate could fill this gap, but a slight noise in the data can be amplified during the high rate re-sampling, which greatly impacts the comparison results. Moreover, the distinct characteristics of the two different sensing modalities and their fidelity differences cause the two domain information even harder to match.
The difficulties of comparing the microphone data with the motion sensor are presented with more details in Appendix~\ref{subsec:difficulty}.

\begin{figure}[t]
	\centering
	\includegraphics[width=1.5in]{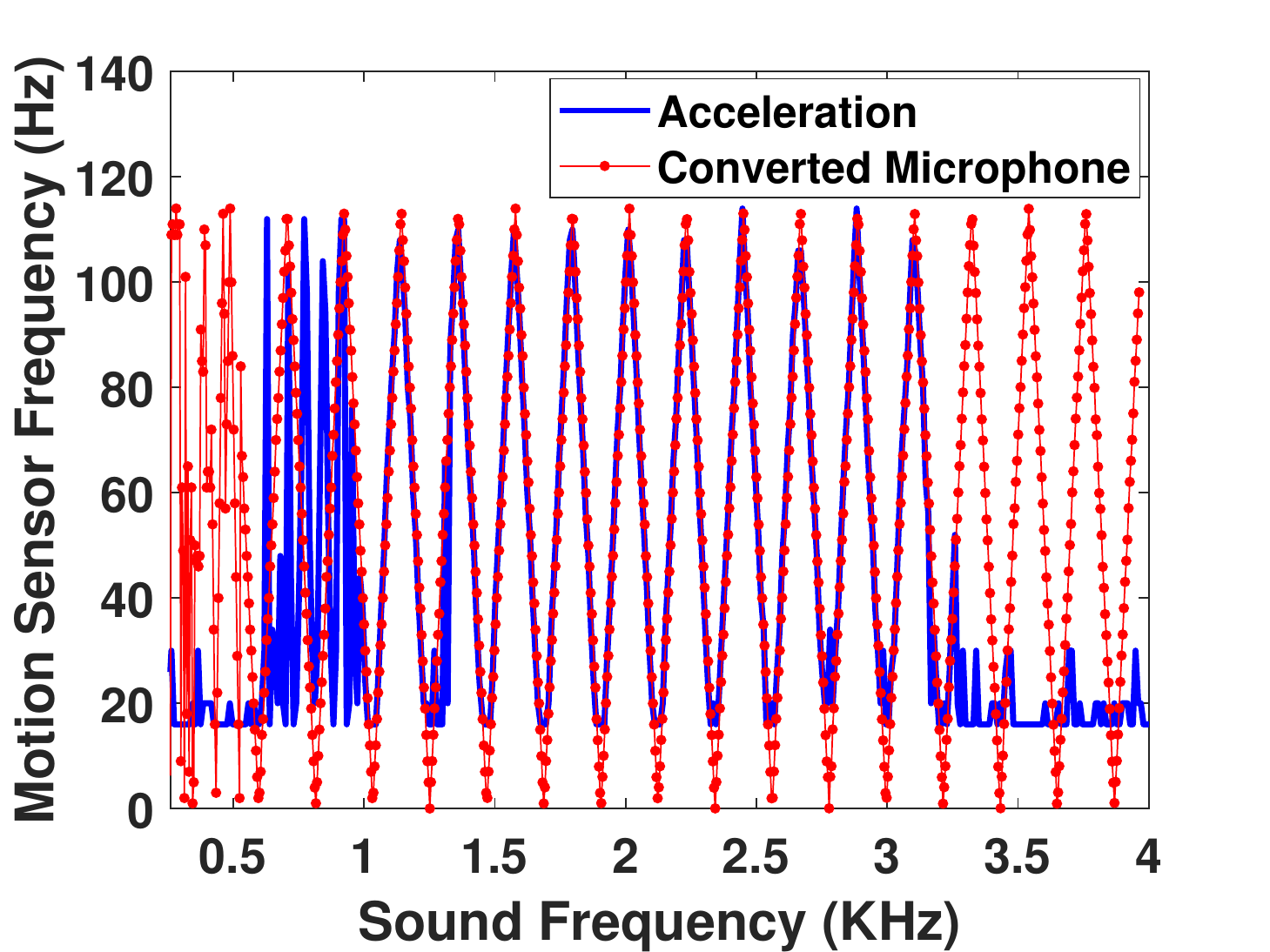}
	\vspace{-4mm}
	\caption{Converting the microphone data of a frequency chirp ($0\sim4KHz$) into the low frequency data based on spectrogram. }
	\label{fig:frequency_transform}
	\vspace{-3mm}
\end{figure}

\section{Prevent Privacy Leakage from Voice Assistant Attacks}

\label{sec:algorithm}


Different from the microphone, the accelerometer shows its unique characteristics when responding to sounds, resulting from its sensor structure, the vibrations of other components in the wearable and the low sensor fidelity. In order to match the voice command sounds captured from the two different sensing modalities to verify the user, our basic idea is to convert the high frequency microphone data into the low frequency data that describes the "equivalent" response of sound on the accelerometer. In this section, we first investigate the complex relationship between the audio domain and the vibration domain. We next introduce our method to leverage such complex relationship to match the microphone sound with the accelerometer readings and provide cross-domain user verification.

\vspace{-2mm}
\subsection{Cross-domain Command Sound Comparison}
\label{subsec:spectrogram}

\textbf{Spectrogram Derivation.}
As the time-domain analysis is shown to be limited to describe the complex relationship between the high frequency microphone and the low frequency motion sensor data (Appendix~\ref{subsec:difficulty}), we resort to the time-frequency analysis and derive the spectrogram (i.e., two-dimensional representation of the signal) to analyze their unique responses to sound. Particularly, we compute the Discrete Time Short Time Fourier Transform (DT-STFT) of the microphone/accelerometer readings $x(n)$ using a sliding window function as expressed in equation~\ref{eq:STFT}.
\begin{equation}
\setlength{\abovedisplayskip}{3pt}
\setlength{\belowdisplayskip}{3pt}
\begin{aligned}
DTSTFT(m, \omega)=\sum_{n=m}^{m+N-1}x(n)w(n-m)e^{-j\omega{n}},
\end{aligned}
\label{eq:STFT}
\end{equation}
where m and $\omega$ are the time index and frequency index of the two dimension signal description, $w(n)$ is a window function, and N is the DT-STFT size of the data in the sliding window (e.g., 2048 for microphone data and 64 for accelerometer readings). 
We then compute the magnitude squared of the DT-STFT $P(m, \omega)=|DTSTFT(m, \omega)|^2$, which is the power spectrum at time $m$. Next, we slide the window by step of size $p$ and obtain the time series of the squared DT-STFT as $S=[P(0, \omega), P(p, \omega),...,P(\frac{M-N}{p}, \omega)]$. Figure~\ref{fig:nonlinearity_chirp} illustrates the derived spectrogram of microphone data and the accelerometer data when responding to a frequency chirp, where each column represents a power spectrum of the microphone/accelerometer when responding to a chirp frequency point.

\begin{figure}[t]
	\begin{center}
		\begin{tabular}{cc}
			\includegraphics[width=1.5in]{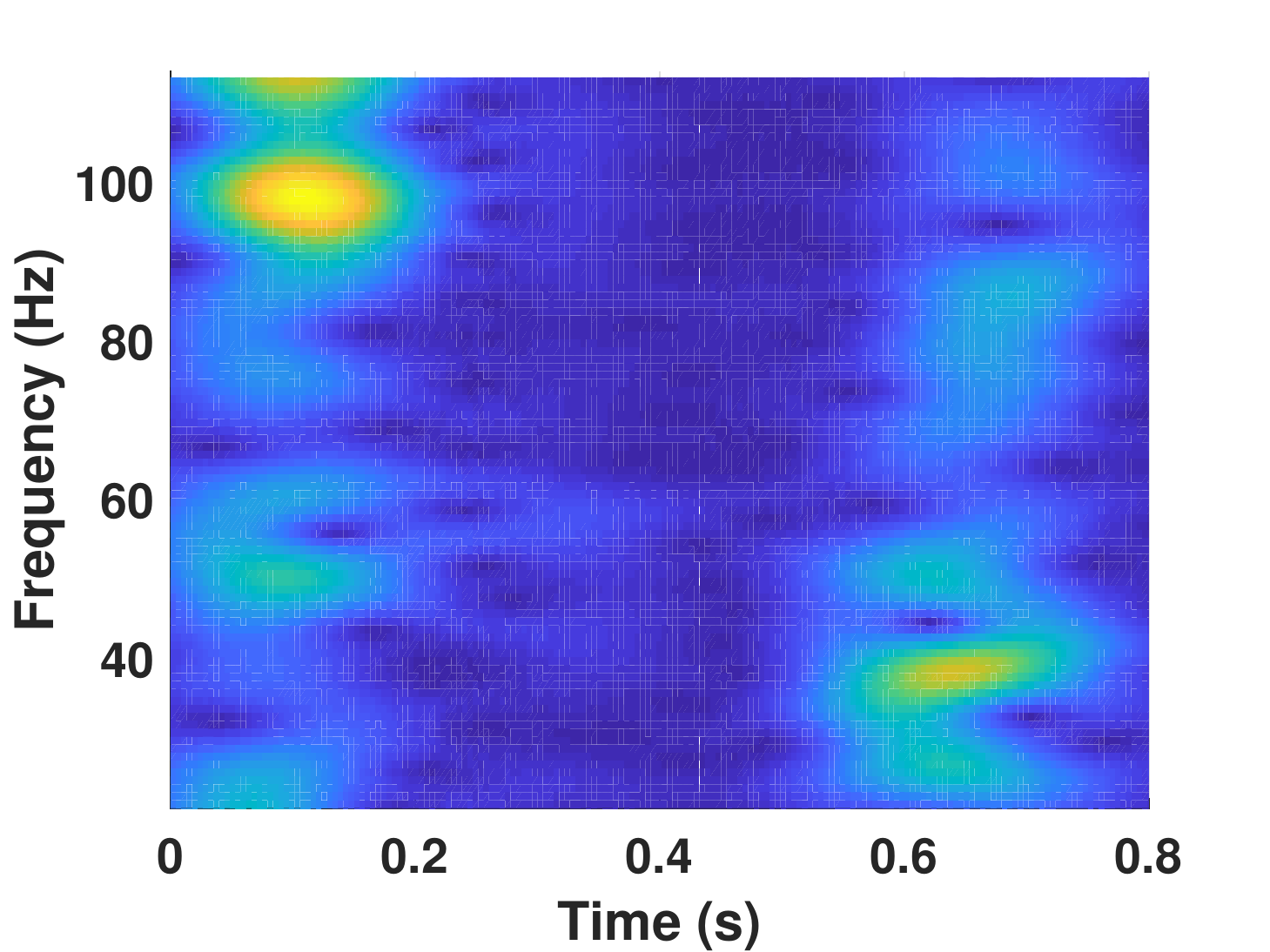}
			&
			\includegraphics[width=1.5in]{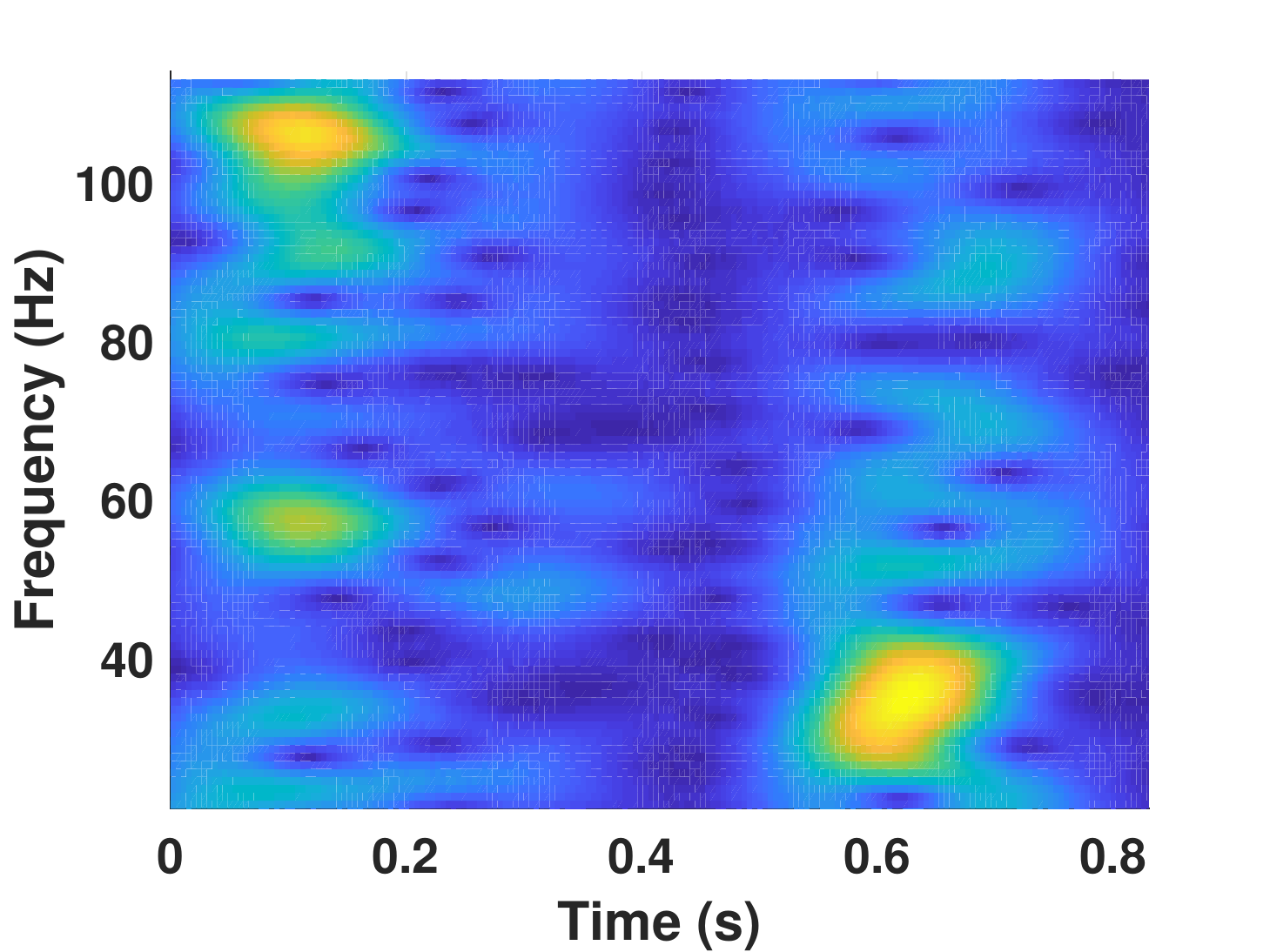}
			
			\\{\scriptsize(a) Motion sensor data} & {\scriptsize(b) Converted Microphone data}\\
		\end{tabular}
	\end{center}
    \vspace{-4mm}
	\caption{Comparison of the motion sensor data spectrogram with the converted microphone data spectrogram (illustrated with the word sound "Alexa"). }
    \vspace{-4mm}
	\label{fig:method_imitate}
\end{figure}

\textbf{Spectrogram-based Frequency Conversion.}
To convert the spectrogram of high-frequency microphone data to the low-frequency one that is comparable to the accelerometer spectrogram, we develop a spectrogram-based frequency conversion method to generate the aliased low frequency microphone data similar to that observed on accelerometer. The high-to-low frequency conversion takes as input original microphone spectrogram point $S_{mic}(t_n,\omega_m)$ and calculates its new position $(t_n,\omega_w)$ in the converted low frequency spectrogram. In particular, the original microphone frequency point $\omega_m$ is mapped to a low frequency point $\omega_w$ based on the signal aliasing equation~\ref{fig:nonlinearity_chirp}, while the time point is unchanged. The resulted new spectrogram is computed as $\hat{S}_{mic}(t_n,\omega_w) = \sum_{n=-\inf}^{\inf}S_{mic}(t_n,win(|\omega_m + n\times \omega_{ws}|))$, where $win()$ is a window function having non-zero value for $[0, \omega_{ws}]$ and $\omega_{ws}$ is the sampling frequency of accelerometer. Based on our spectrogram-based frequency conversion, the spectrogram of the microphone data can be transformed into the low-frequency spectrum. For example, for the single frequency chirp, the spectrogram in Figure~\ref{fig:nonlinearity_chirp}(a) can be transformed to the similar aliased ``Zigzag'' form as the accelerometer's spectrogram in Figure~\ref{fig:nonlinearity_chirp}(b). But the human sound is more complicated than a single frequency chirp. To convert the microphone spectrogram to the more ``precise'' low frequency form as the accelerometer, we also need to study the unique characteristics of accelerometer.



\textbf{Frequency and Amplitude Selection.}
Except the low sampling rate, the accelerometer's sensor structure, the vibrations of other electric components in the wearable device and the low sensor fidelity all cause the accelerometer to respond to sounds differently from a microphone. In particular, these factors cause the accelerometer to subdue some frequencies while in favor of responding to other frequencies with higher amplitude. Moreover, being not dedicated for recording sound, the accelerometer shows lower sensitivity to sounds and some small volume sounds may not result in readings in the accelerometer but can be easily recorded by microphones.
To reveal the complex relationship between the two sensing modalities when recording sounds, we compare the accelerometer's spectrogram with the converted low-frequency spectrogram of microphone for every frequency point of a chirp signal. In particular, we extract the maximum amplitude on each spectrum (i.e., column of the spectrum) at every time index and obtain a clear \textit{frequency sweeping curve} for the two sensing modalities as shown in Figure~\ref{fig:frequency_transform}. Compared with the microphone, the accelerometer only respond to a small frequency range ($700$Hz - $3300$Hz) (i.e., blue zigzag frequency sweeping curve). Moreover, for this frequency range, the frequency sweeping curves of the two sensing modalities match well. Besides the response frequency range, we also analyze the amplitude of the sound that could generate responses on the wearable's accelerometer. In particular, we find that when the sound is greater than $70$dB, the sound can leave obvious responses on the wearable, which is consistent with the observations in Accelword~\cite{zhang2015accelword}. Therefore, in order to facilitate the similarity comparison between the two sensing modalities, wearID needs to convert the microphone data to low frequency data following the same frequency range and response amplitude of the accelerometer, which maximize their acoustic response intersections. Figure~\ref{fig:method_imitate}(b) illustrates an example of the converted spectrogram of the microphone data for the word sound "Alexa", which shows an "equivalent" low frequency spectrogram as that of the accelerometer in Figure~\ref{fig:method_imitate} (a).

\textbf{Spectrogram-based Conversion Algorithm.}
We now introduce the spectrogram-based microphone data conversion algorithm, which integrates the spectrogram-based frequency conversion, the frequency selection and the amplitude selection. As shown in Appendix Algorithm~\ref{al:spectrogram_transformation}, the transformation algorithm takes microphone spectrogram $S_{mic}$ and the sampling rate of the accelerometer $f_{ws}$ as input and calculates the new spectrogram $\hat{S_{mic}}$ that locates within the low-frequency range (e.g., $0 - 200Hz$) as the output. For each column of spectrogram (i.e., $1-T$), the algorithm only selects the power spectrum point within frequency $700Hz$ to $3300Hz$ and with the magnitudes greater than $70$dB for conversion. Spectrogram-based frequency conversion is then performed to the selected spectrogram points based on equation~\ref{eq:nonlinearity}. If multiple spectrogram points are mapped to the same point in the new spectrogram, their magnitudes are added together. Finally, the converted microphone spectrogram is in the same frequency range of accelerometer and could describe the sound in vibration domain similar to the accelerometer.

\begin{figure}[t]
	\centering
	\begin{tabular}{cc}
		\includegraphics[width=1.5in]{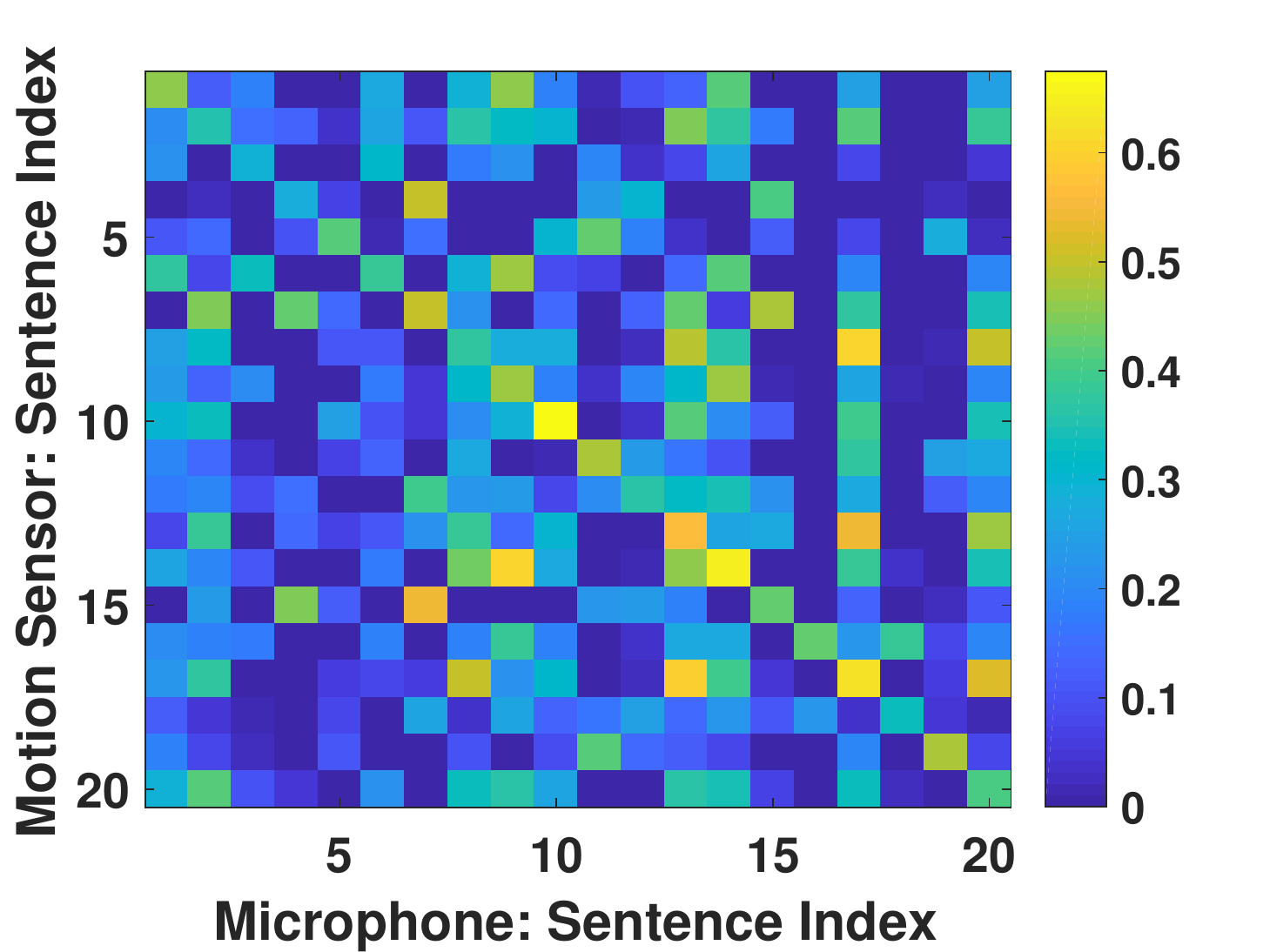}
		&
		\includegraphics[width=1.5in]{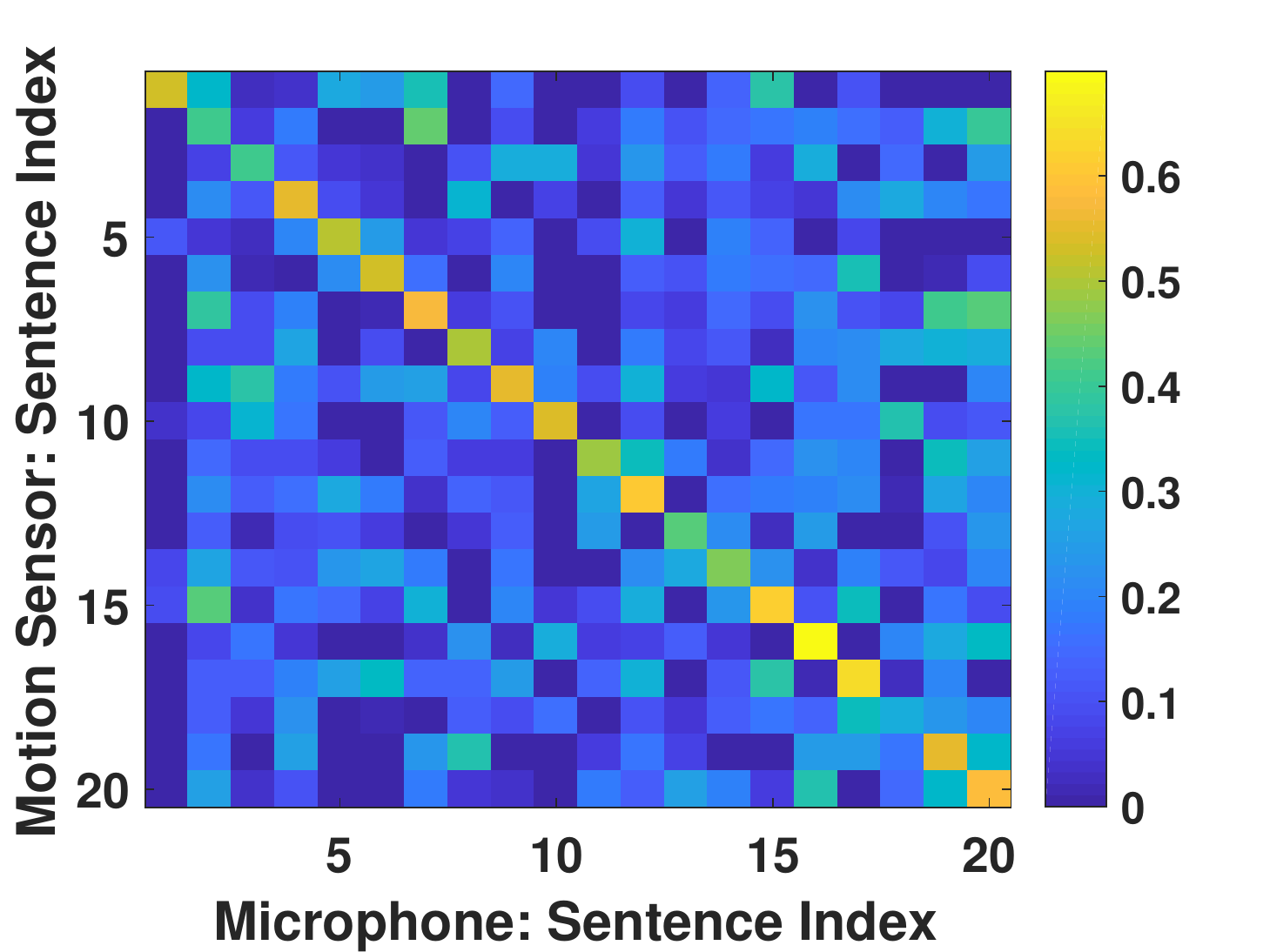}
		
		\\{\scriptsize(a) Void sound of words} & {\scriptsize(b) Voice sound of sentences}\\
	\end{tabular}
	\vspace{-4mm}
	\caption{The spectrogram correlation based on our method.}
	\vspace{-5mm}
	\label{fig:correlation_map_word}
\end{figure}

\vspace{-2mm}
\subsection{Legitimate User Verification}
\textbf{Spectrogram Normalization.}
The scales of amplitude are greatly different in accelerometer and microphone readings. Before checking the similarity of the two modalities' data, we develop the 2D-interpolation scheme and the 2D-normalization scheme to normalize the two spectrograms both in time and magnitude. In particular, the 2D-interpolation scheme performs row-based interpolation to align the two spectrograms. The 2D-normalization resolves the scale differences of the two spectrograms' magnitudes and conduct column-based normalization to unify the magnitudes within $[0,1]$. The 2D-normalization is described in equation~\ref{eq:normalization}:
\begin{equation}
\setlength{\abovedisplayskip}{3pt}
\setlength{\belowdisplayskip}{3pt}
\begin{aligned}
{S}_{norm}(t_n,w_m)=\frac{S(t_n,w_m)-S_{min}(t_n)}{S_{max}(t_n)-S_{min}(t_n)},
\end{aligned}
\label{eq:normalization}
\end{equation}
$S(t_n,w_m)$ is a power spectrum point at time $t_n$ and $S_{max}(t_n), S_{min}(t_n)$ denote the maximum and minimum power spectrum magnitude at time $t_n$.

\textbf{Cross-domain Similarity Calculation based on Shift 2D-Correlation.}
Next, we match the command sounds across the audio domain and the vibration domain and calculate the 2D-correlation coefficient between the microphone and accelerometer spectrograms using equation:
$Corr(S_{mic}, S_{acc})=\frac{A\times{B}}{\sqrt{A^{2}\times{B^2}}}$, 

where A, B represent two spectrogram matrixes. Note that the microphone data and the accelerometer data are coarsely synchronized in~\ref{subsubsec:synchronization} and the synchronization error may affect the similarity comparison accuracy. To address this issue, we conduct the \textit{Shift 2D-Correlation} when computing the correlation between the spectrograms of the microphone and accelerometer. In particular, we fix the microphone spectrogram and shift the spectrogram of accelerometer one index by one index along time axis to calculate the similarity. More specifically, we use a sliding window with a fixed size and shift it to left or right on the accelerometer's spectrogram within time $T$ (e.g., $500ms$). The 2D-correlation is calculated for each shift and the maximum 2D-correlation coefficient is found as the similarity score of the two domain information. A threshold-based method is then applied to examine the similarity score and verify the user.
Figure~\ref{fig:correlation_map_word}(a) illustrates the similarity comparison result of the above method to differentiate 20 different words. Clearly, the diagonal comparisons show much higher correlation coefficients, which can be distinguished from that of the different words. Figure~\ref{fig:correlation_map_word}(b) further confirms the efficiency of our method to differentiate a user's 20 voice commands (i.e., sentences), which are distinguished better, because sentences contains much more voice information than single words. Compared with Figure~\ref{fig:downsample_timecorrelation}, it is clear that our method can verify whether the recorded sounds in the two domains are from the same sound source, because our method identifies the complex relationship between the two domains and captures the most response intersections between them.

\vspace{-2mm}
\subsection{Data Preprocessing}




\begin{figure}[t]
	\centering
	\includegraphics[width=3in]{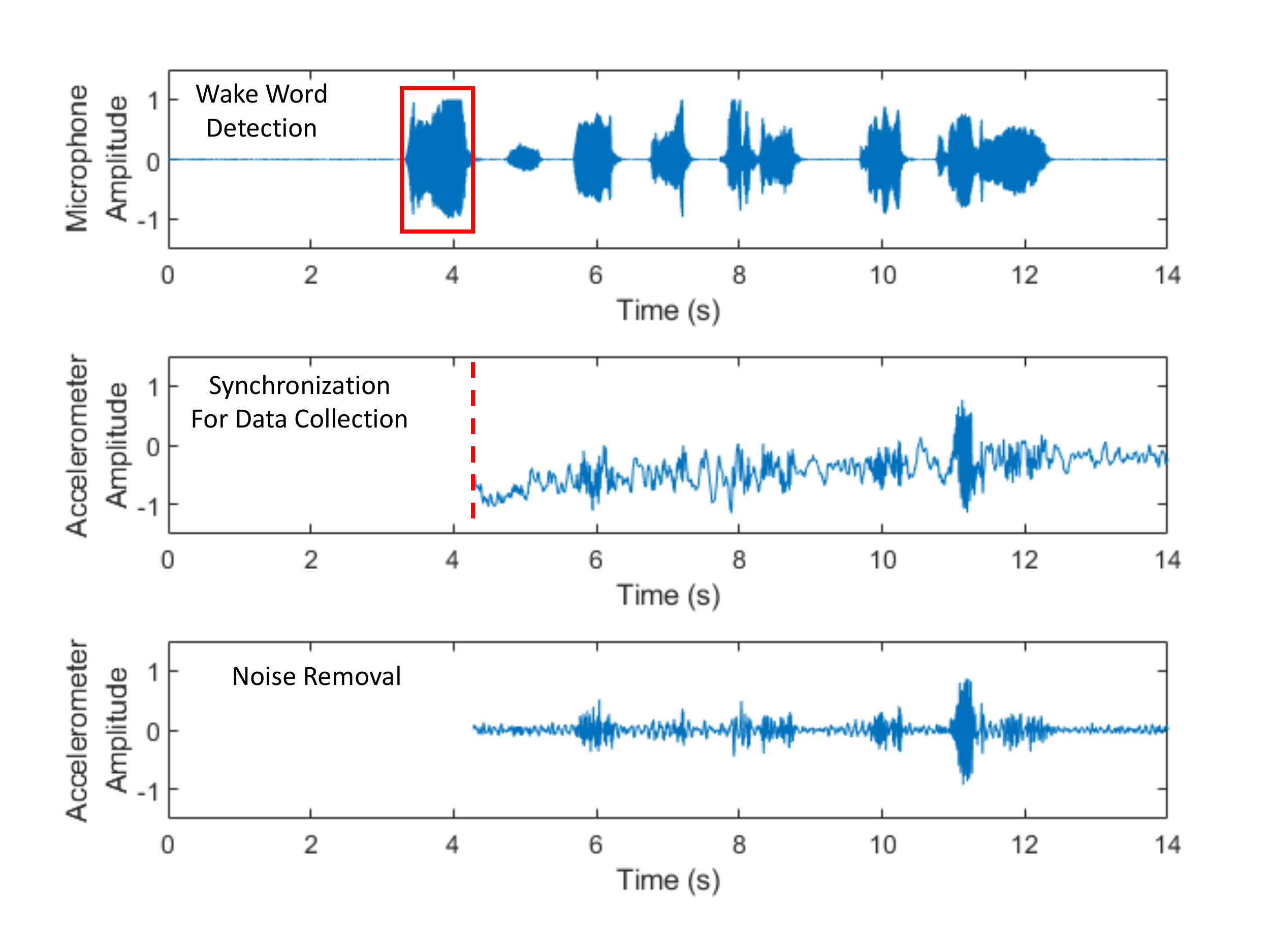}
	\vspace{-4mm}
	\caption{Synchronization of the microphone data (8000Hz) and accelerometer data (200Hz) and the hand vibration noise removal from accelerometer data. }
	\label{fig:synchronization}
	\vspace{-5mm}
\end{figure}

\subsubsection{Coarse-grained Synchronization}
\label{subsubsec:synchronization}
Coarse-grained synchronization triggers the VA device and wearable to collect the command sound and perform coarse synchronize for both devices. The existing VA system requires the user to speak a wake word such as "OK Google" and "Alexa" to wake up the VA device before taking any voice commands. WearID integrates such method to trigger the verification process and start the data collection on both microphone and accelerometer. In particular, we develop two alternative approaches, the WiFi communication-based method and the parallel wake-word detection method. 1) WiFi communication-based method leverages the existing setting that the wearable device and the VA device are connected to the same WiFi network~\cite{androidwear2018}. The VA device after being waked sends a message through connected WiFi network to the wearable device to trigger its data collection. While standalone wearables directly receive the WiFi packet, the wearables that work with a paired smartphone can receive the message relayed by the smartphone's Bluetooth. In addition, the time lag between the microphone and accelerometer data is usually less than $40$ms, which is mainly caused by the network delay and the system time differences. Figure~\ref{fig:synchronization}(a) and (b) show the accelerometer readings of the WiFi communication-based method, where the accelerometer starts recording after receiving the waked VA device's message. Note that the data on the microphone and the accelerometer are roughly synchronized. 2) As an alternative method, the parallel wake-word detection method requires the wearable to detect the wake word in parallel with the VA device to trigger command sound recording. The wearable reuses the accelerometer data from an ongoing fitness APP to recognize the wake word. Our study shows that a wake word can be recognized based on accelerometer from $10$ words with $83\%$ accuracy by Random Forest. Moreover, the wake-word detection on the wearable can be further improved if assisted with a hand motion detection scheme (e.g., a lift-hand motion).

\vspace{-1mm}
\subsubsection{Noise Removal}
We note that the accelerometer data on the wearable contains much vibration noises (e.g., hand vibrations) as shown in Figure~\ref{fig:synchronization}(b). While these vibration noises are in low-frequency compared to acoustic signals, we apply a high-pass filter (e.g., cutoff frequency $30Hz$) to remove the noise and obtain the accelerometer data that describes the command sound more precisely as shown in Figure~\ref{fig:synchronization}(c). We find that the high-pass filter can effectively reduce the noise introduced by hand movements or other mechanical vibrations and the resulted accelerometer data (Figure~\ref{fig:synchronization}(c)) shows a slightly similar shape to the microphone data (Figure~\ref{fig:synchronization}(a)). To obtain the microphone data that precisely captures command sound and remove the acoustic noises, we apply a bandpass filter (e.g., $300 - 4000Hz$) to filter out the acoustic sounds beyond human voice frequency range.

\vspace{-1mm}
\subsubsection{Command Sound Segmentation}
We next search for the \textit{starting point} and \textit{ending point} of the voice command on both the microphone and the accelerometer data to identify the voice command segments respectively. In particular, we analyze the moving variance of the data amplitudes and extract the envelope that covers the command sound. We then apply a threshold-based method to search for the starting/ending points. The command sound segmentation on microphone is easy and accurate. But on the accelerometer, the segmentation errors may be large due to its low sensitivity to sound and low sampling rate. To address this issue, we apply the microphone-based command sound segmentation results to assist the command sound segmentation on the accelerometer. Because both data are coarsely synchronized, we search for the starting point on accelerometer within a window $W_{T}$ after the starting point identified on the microphone data. Moreover, we calculate the ending point on accelerometer data based on the command sound length obtained from the microphone data.

\begin{figure}[t]
	\begin{center}
		\begin{tabular}{cc}
			\includegraphics[height=1.0in]{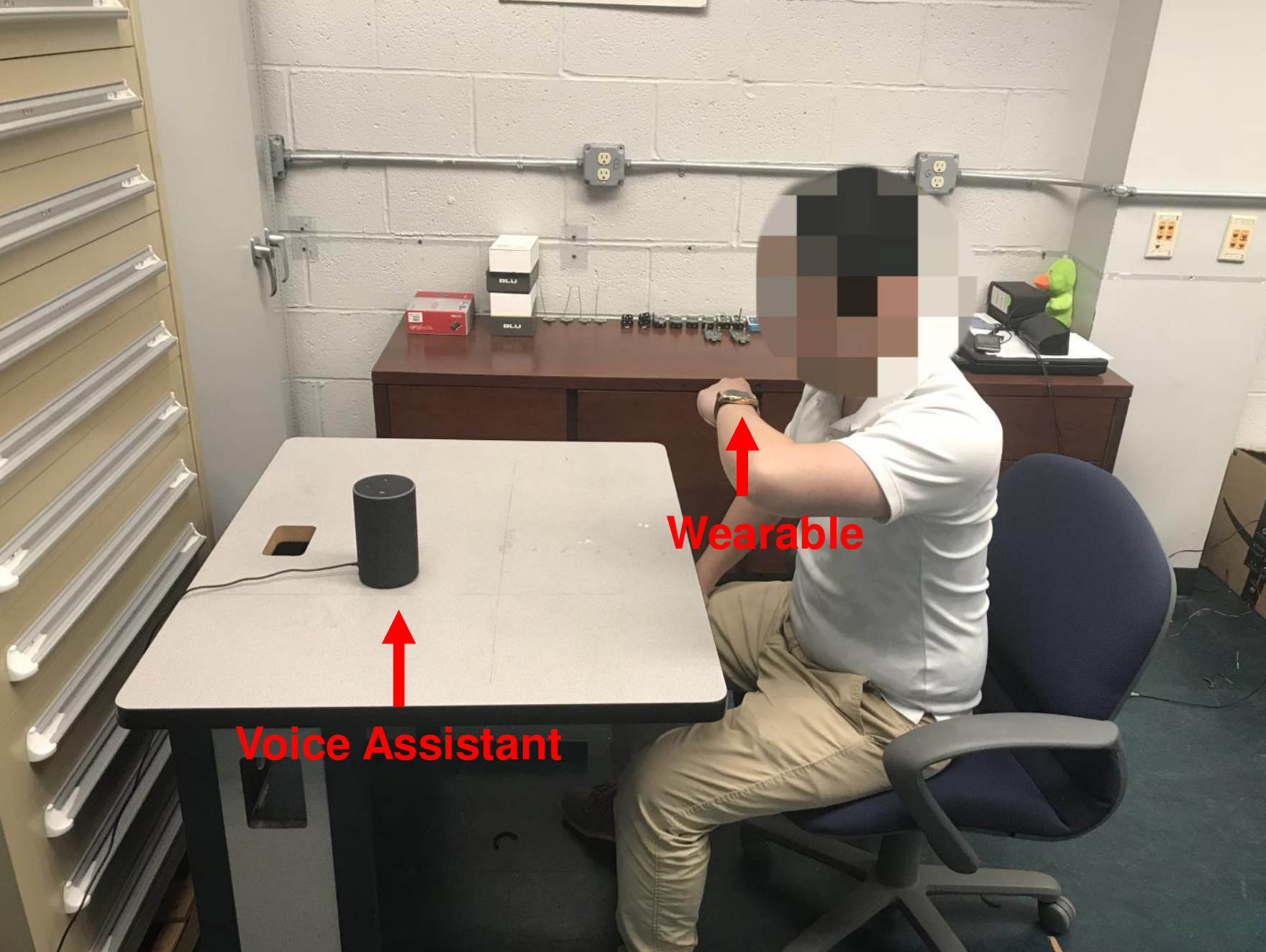}
			&
			\includegraphics[height=1.0in]{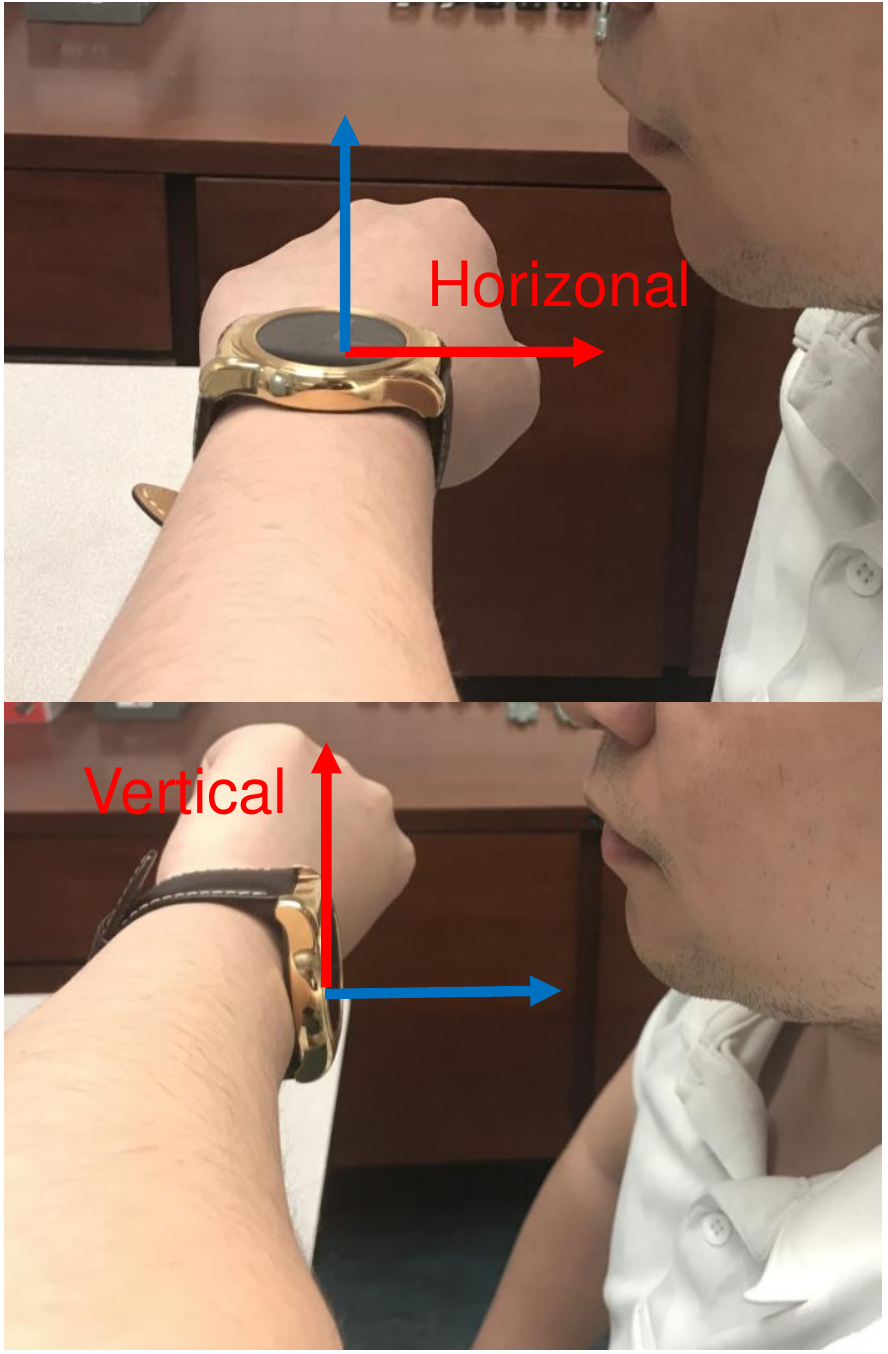}
			\\{\scriptsize(a) Experiment Setup} & {\scriptsize(b) Horizontal and Vertical}
		\end{tabular}
	\end{center}
	\vspace{-4mm}
	\caption{The experimental setup and the two representative ways of holding the wearable device. }
	\vspace{-5mm}
	\label{fig:setup}
\end{figure}

\vspace{-1mm}
\section{Performance Evaluation}
\label{sec:experiment}

\vspace{-1mm}
\subsection{Experimental Methodology}




\textbf{Device.} To evaluate WearID, two smartwatch models, Huawei 2 sport ($100$Hz) and LG W150 ($200$Hz) are involved to collect accelerometer readings. The accelerometer specifications of the two wearable devices are listed in Appendix Table~\ref{table:devices}. The two smartwatches run Android Wear OS $2.0$ with Bluetooth LE. Although the sampling rate of the accelerometers could reach $4000$Hz, the vendors constrain the sampling frequencies to be under $200$Hz to ensure low power consumption. We record the voice commands leveraging a typical VA device, Google Home, which supports $8\sim96kHz$ audio recording.
We use Logitech S120 speaker~\cite{logitechs120} to conduct replay attacks and hidden voice commands. To evaluate WearID against ultrasound attack, we use a function generator (i.e., Keysight Technologies 33509B~\cite{Keysight33509B}) together with a tweeter speaker~\cite{tweeterspeaker} which has the $2kHz\sim25kHz$ frequency range.



\textbf{Experimental Setup.} We evaluate the performance of WearID to verify VA system users in a typical office environment, where regular ambient noises (e.g., air condition and people walking around) are presented. The experimental setup is shown in Figure~\ref{fig:setup}. The participant speaks voice command to a VA system placed 1 meter way while holding the wearable device horizontally or vertically to mouth. Note that the users can choose to hold the wearable in any way, and without loss of generality, we evaluate WearID with the two typical ways of using the wearable (e.g., checking time). To imitate the hidden voice commands and the ultrasound attack, we use the experimental setup as shown in Figure~\ref{fig:preliminary_setup}. In particular, we examine an extreme case where the loud speaker and the tweeter speaker are placed at $25cm$ distance to the wearable. In practical attack, the speakers could not be placed too close to the user without causing the user's attention.



\textbf{Data Collection.} 
We involve 10 participants to test WearID under the normal situation and various attacks over a six-month period. The participants are asked to speak $20$ representative voice command sentences as listed in Appendix Table~\ref{tb:privacyleakage} while wearing different wearables and holding it in two typical ways. From each participant, $80$ voice command sound samples are collected. Besides, $10$ types and $100$ hidden voice command sounds are utilized to evaluate WearID against hidden voice command attack~\cite{hiddenvoice}, and a frequency sweeping signal from $15kHz\sim25kHz$ is used to evaluate WearID against ultrasound attack. In total, $1000$ data samples are collected for microphone and motion sensor respectively.

\textbf{Evaluation Metrics.} To evaluate our system performance, we define the following five different metrics: true positive rate (TPR) is the percentage of legitimate voice commands being correctly verified; false positive rate (FPR) is the percentage of the adversaries' voice commands that pass the verification system; receiver operating characteristics (ROC) curve is generated by plotting the TPR against the FPR under thresholds from $0$ to $1$ with a step of $0.01$; area under the ROC curve (AUC); False Negative Rate (FNR) equaling to $1 - TPR$ is the percentage of legitimate users' commands being incorrectly rejected.

\begin{figure}[t]
	\begin{center}
		\begin{tabular}{cc}
			\includegraphics[width=1.5in]{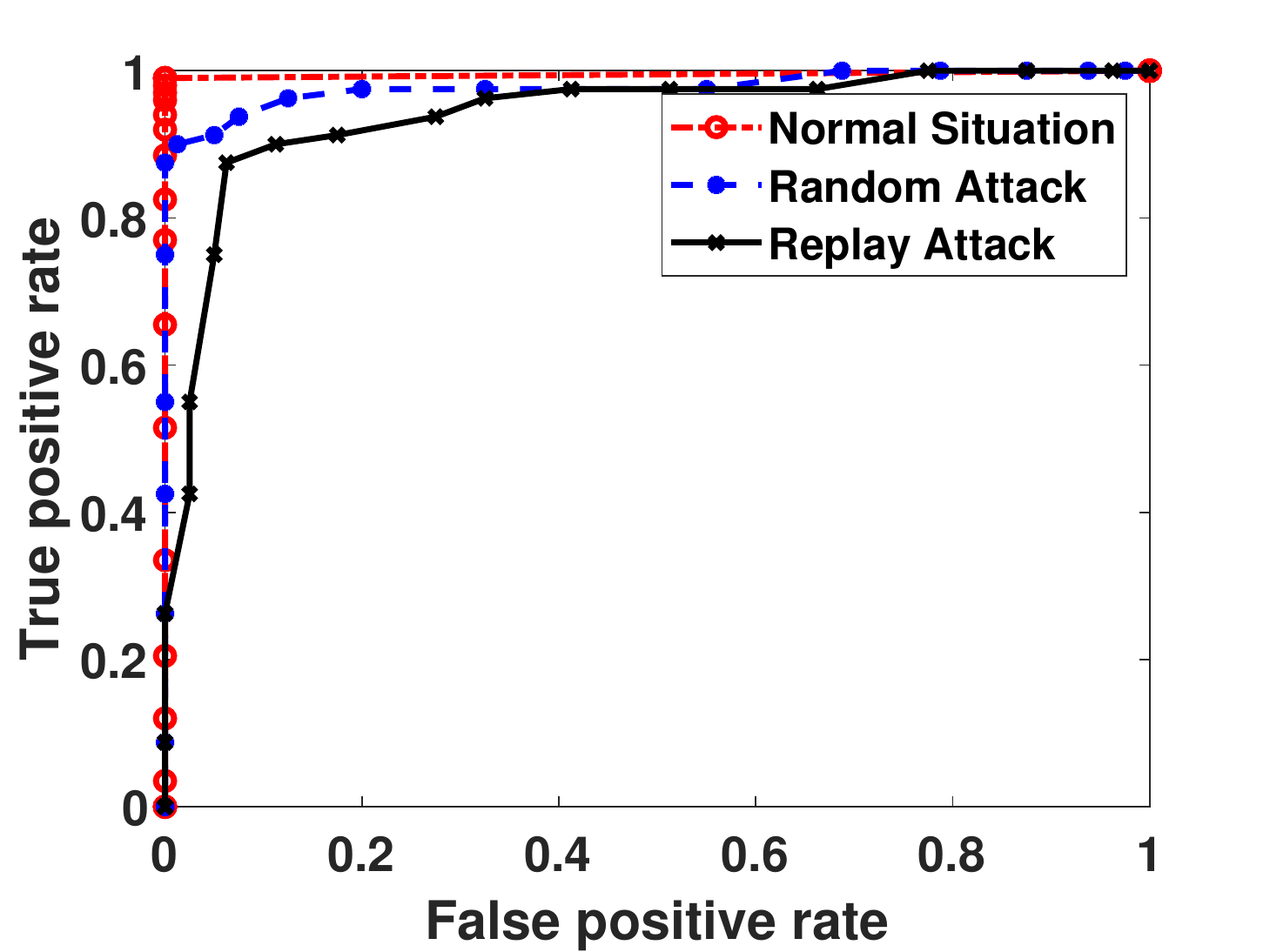}
			&
			\includegraphics[width=1.5in]{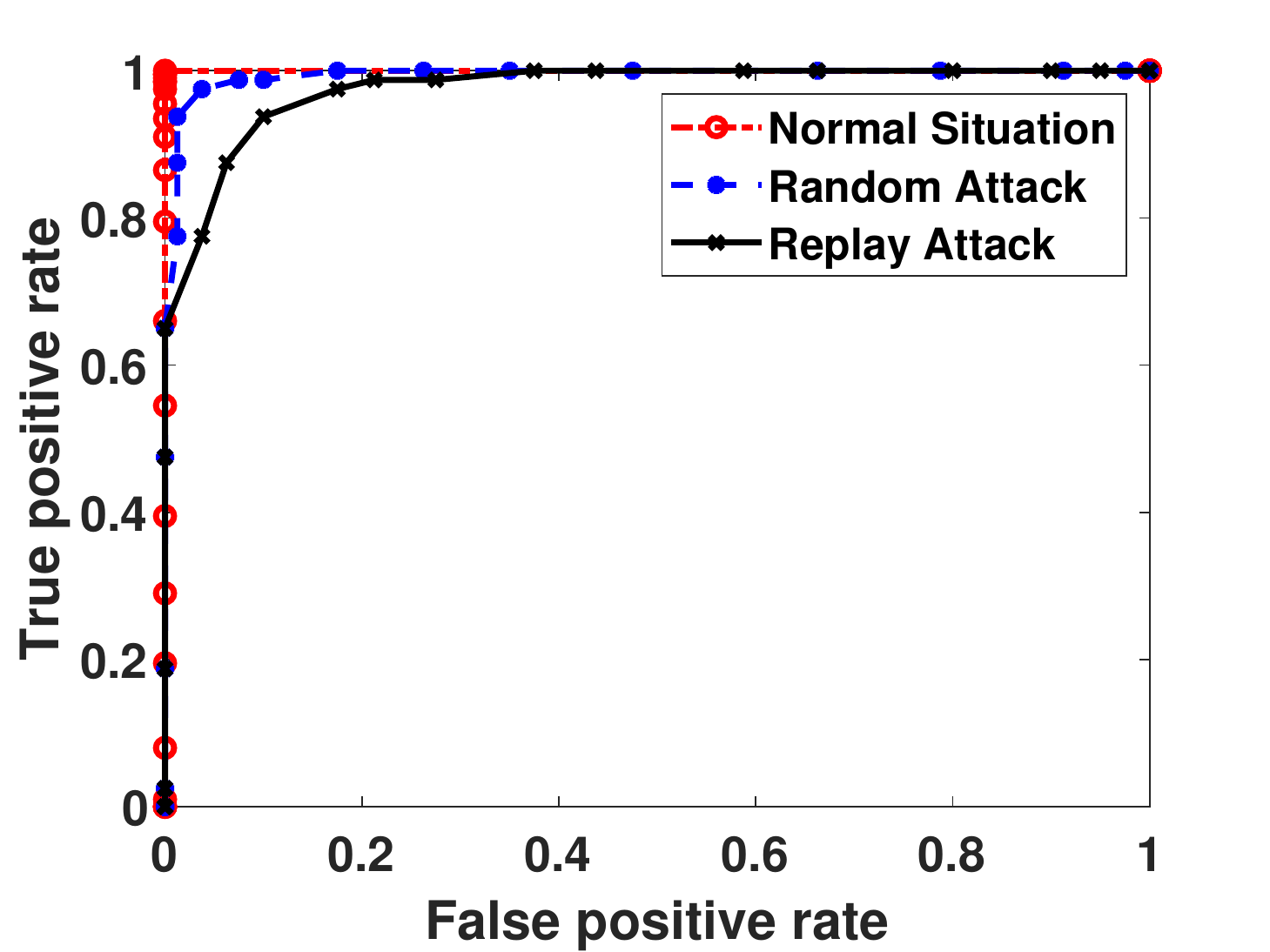}
			\\{\scriptsize(a) Horizontal} & {\scriptsize(b) Vertical}
		\end{tabular}
	\end{center}
	\vspace{-4mm}
	\caption{Average ROC curve of verifying the user using Huawei watch 2 under normal situation, random attack and impersonate/replay attacks.}
	\label{fig:huawei2_roc2}
	\vspace{-5mm}
\end{figure}

\begin{figure}[t]
	\begin{center}
		\begin{tabular}{cc}
			\includegraphics[width=1.5in]{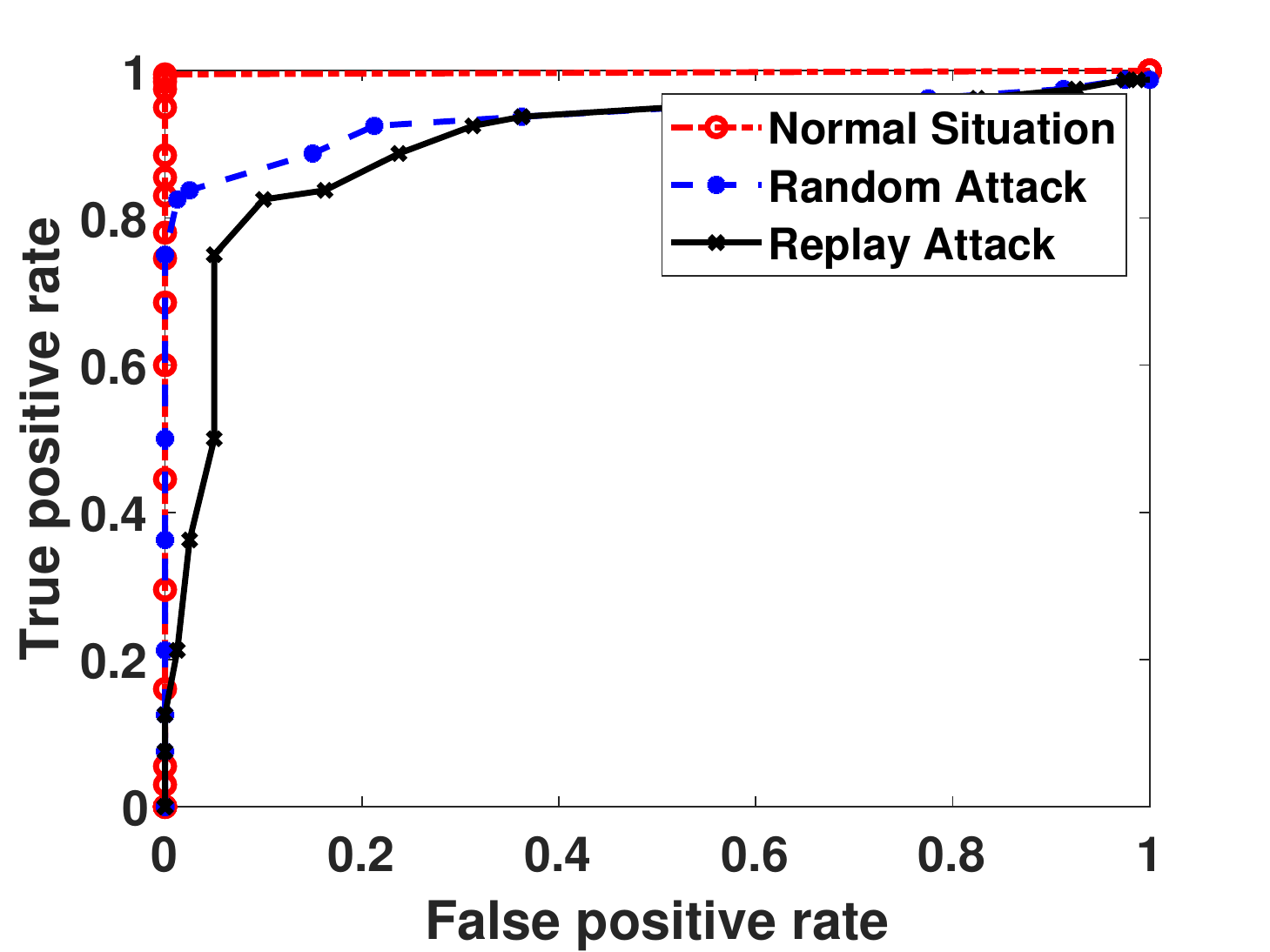}
			&
			\includegraphics[width=1.5in]{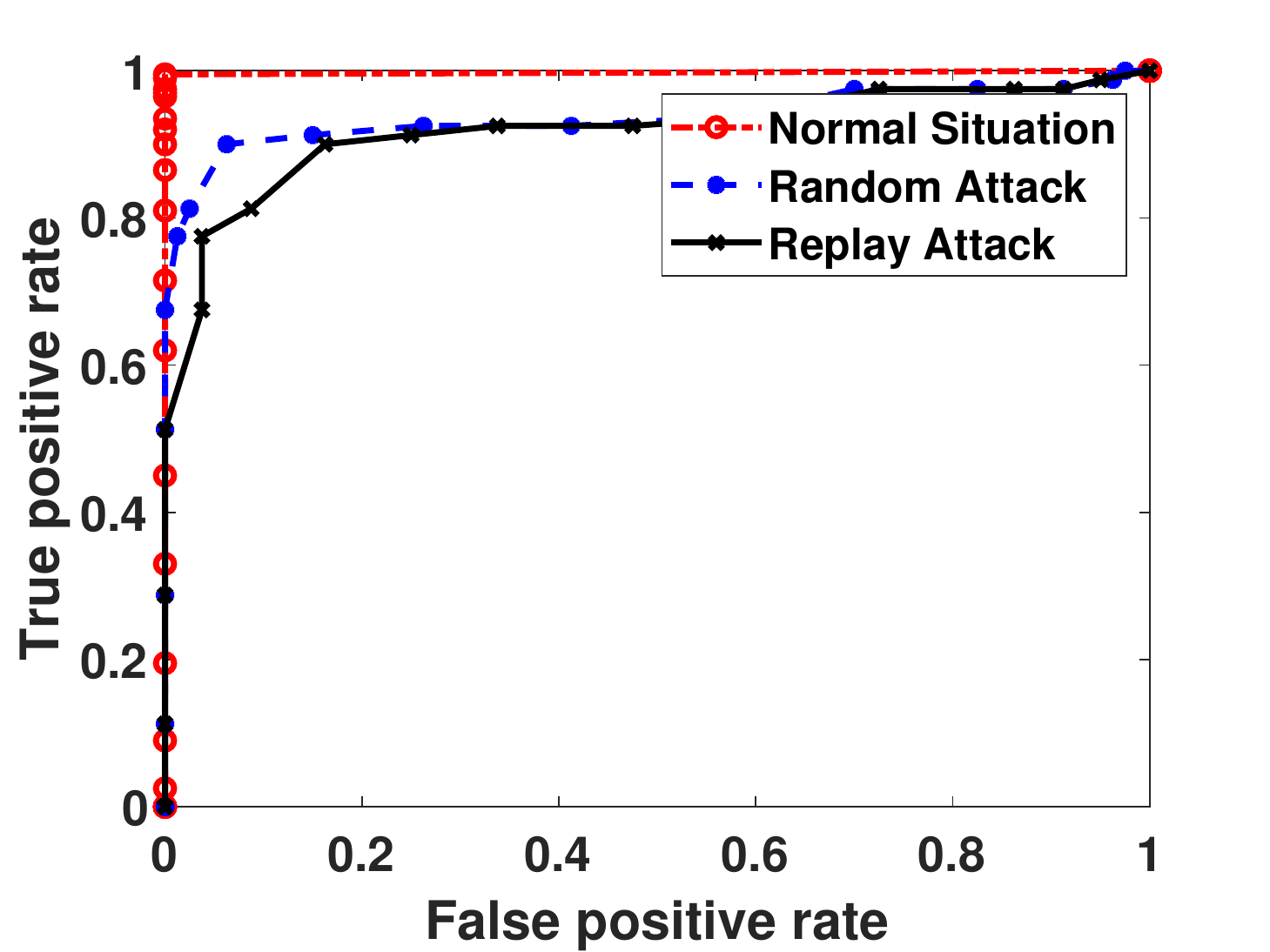}
			\\{\scriptsize(a) Horizontal} & {\scriptsize(b) Vertical}
		\end{tabular}
	\end{center}
	\vspace{-4mm}
	\caption{Average ROC curve of verifying the user using LG Urban W150 under normal situation, random attack and impersonate/replay attacks.}
	\label{fig:w150_roc2}
	\vspace{-5mm}
\end{figure}

\vspace{-2mm}
\subsection{Normal Situation}
We first evaluate WearID in the normal situation when the user accesses the VA device while wearing a wearable and there is no targeted attack. The VA device records the user's voice command sound and the wearable device records the same sound simultaneously, if it is worn by the user. But if the wearable is not worn by the user, it only records the environmental noises (e.g., acoustic noises and mechanical noises), because the wearable is not presented. The red curves in Figure~\ref{fig:huawei2_roc2}(a) and (b) present the ROC curve of WearID to verify the legitimate users using Huawei watch 2 under the normal situation, when the user uses the wearable in two typical ways. In particular, WearID achieves $99.8\%$ TPR and $0\%$ FPR to recognize the legitimate users with Huawei watch 2 for both holding ways. The red curves in Figure~\ref{fig:w150_roc2}(a) and (b) shows the ROC obtained by LG W150 smartwatch. We find that WearID recognizes $99.6\%$ legitimate users' commands for both holding ways while FPR is $0\%$. The false negative rate in the normal situation for both smartwatches is $0.02\% ~ 0.04\%$, which indicates that WearID is robust and accurate to support the users' daily usage of the VA device.
\vspace{-2mm}
\subsection{Attack on User's Absence}
When the user is not present to the VA system, an adversary can reach to the VA device and perform random attack or more sophisticated impersonate and replay attacks to access the legitimate user's privacy in the VA device. During such attacks, the VA device's microphone picks up the attacking sound and the legitimate user's wearable, while in a different place with the user, may record the owner's sound and the environmental noises.
\vspace{-2mm}
\subsection{Differentiating People's Voices. }
\begin{figure}[t]
	\begin{center}
		\begin{tabular}{cc}
			\includegraphics[width=1.6in]{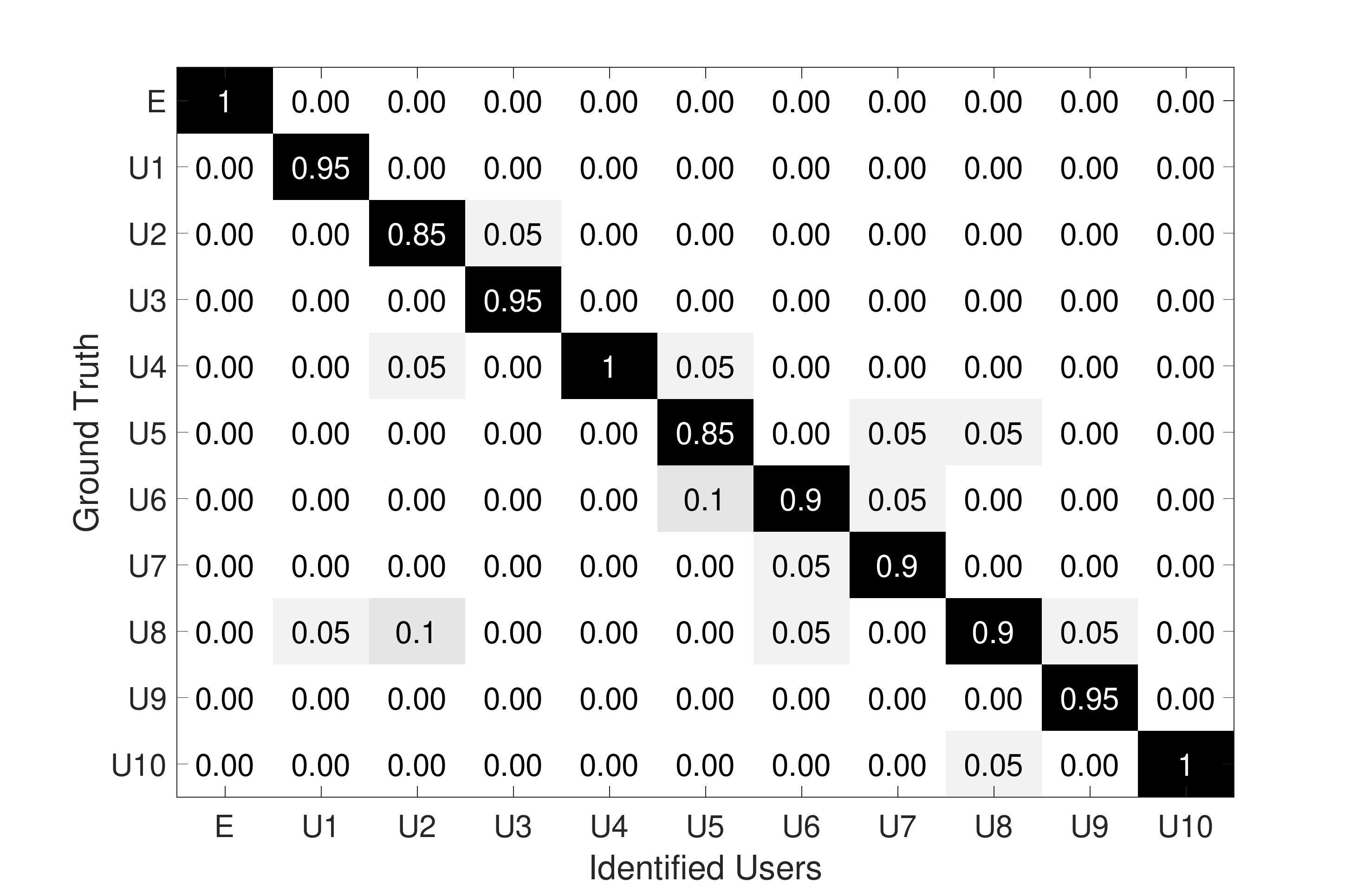}
			&
			\includegraphics[width=1.6in]{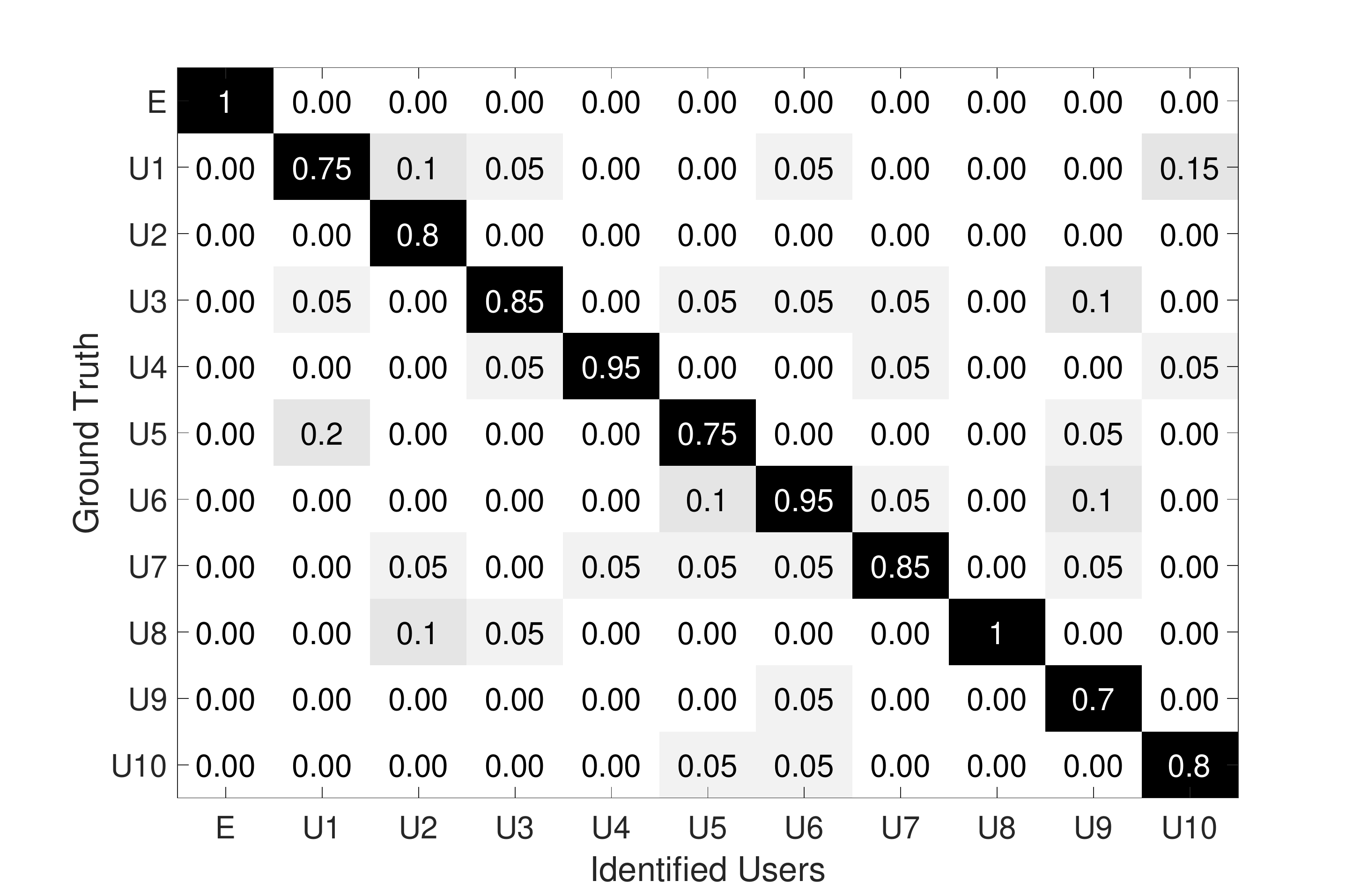}
			\\{\scriptsize(a) Huawei watch 2} & {\scriptsize(b) LG Urban W150}
		\end{tabular}
	\end{center}
	\vspace{-4mm}
	\caption{Confusion matrix for distinguishing people's voices based on the same commands with horizontal holding way.}
	\vspace{-5mm}
	\label{fig:conf_horizontal}
\end{figure}

\begin{figure}[t]
	\begin{center}
		\begin{tabular}{cc}
			\includegraphics[width=1.6in]{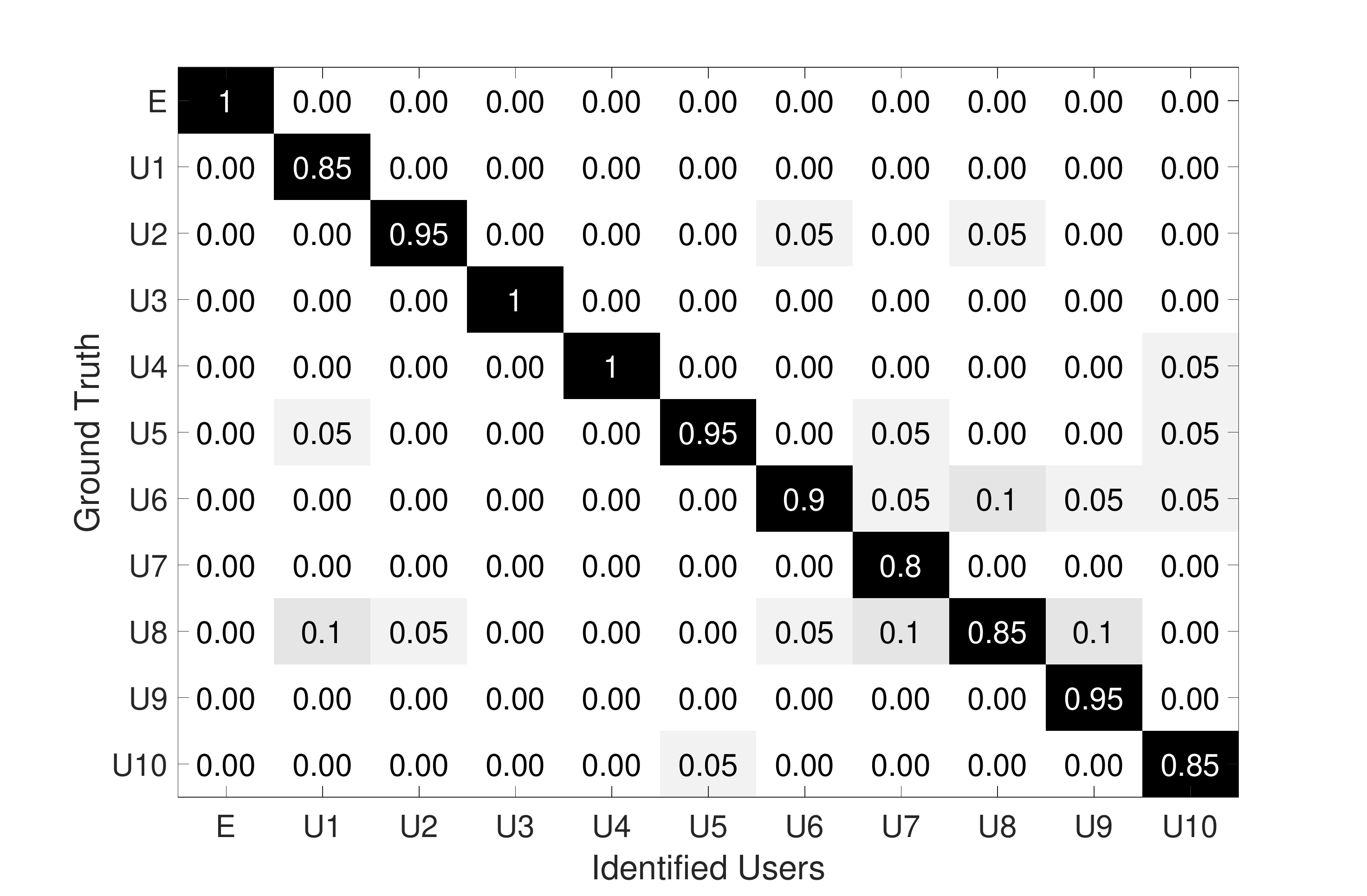}
			&
			\includegraphics[width=1.6in]{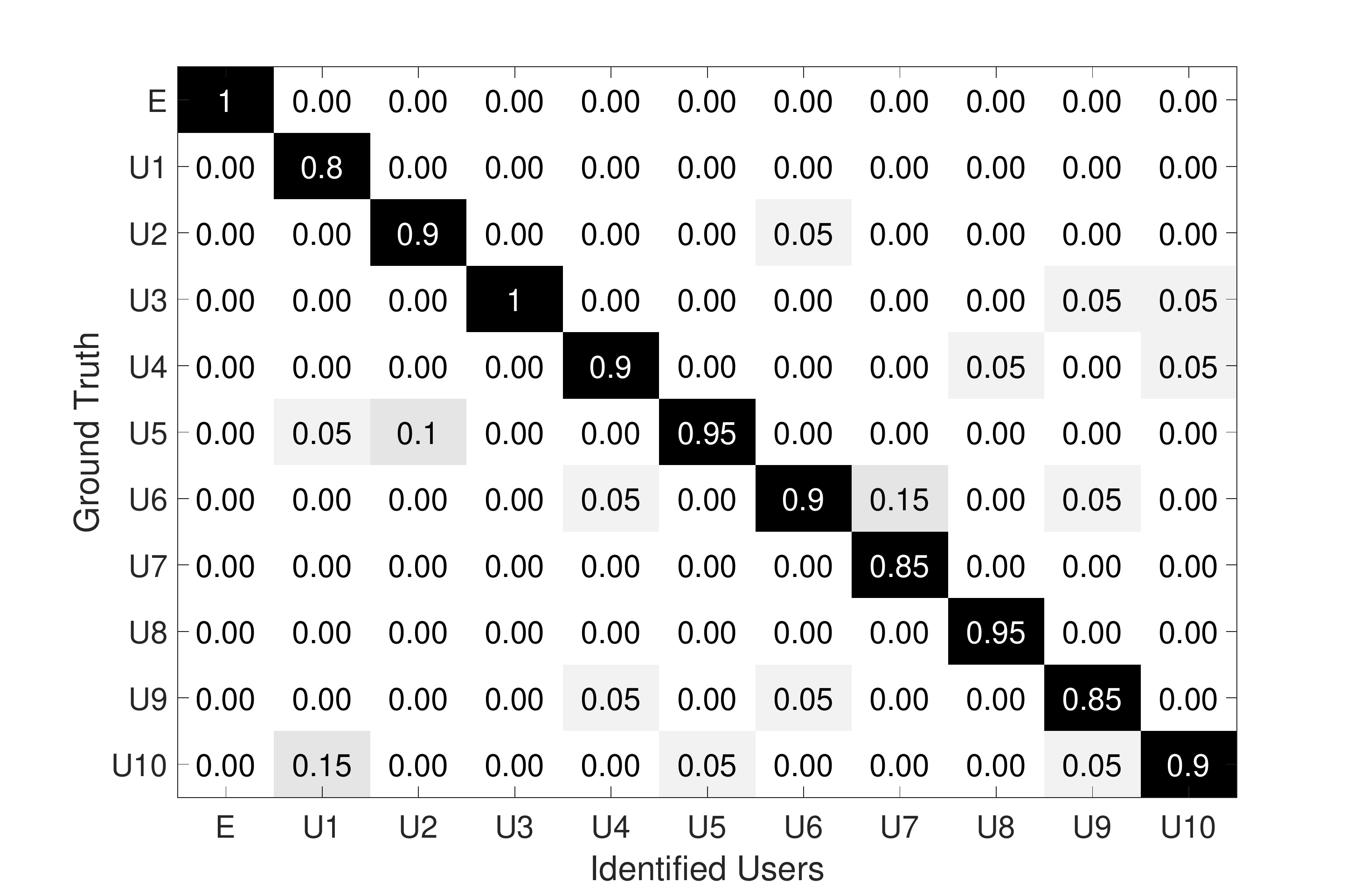}
			\\{\scriptsize(a) Huawei watch 2} & {\scriptsize(b) LG Urban W150}
		\end{tabular}
	\end{center}
	\vspace{-4mm}
	\caption{Confusion matrix for distinguishing people's voices based on the same commands with vertical holding way.}
	\vspace{-7mm}
	\label{fig:conf_vertical}
\end{figure}

We first evaluate WearID's capability to differentiate people's command sounds across two domains, because when the user is absent with his/her wearable, the voice sounds received by the VA device may be different from that on the wearable. We consider an extreme case where we ask participants to speak the same voice commands to either the VA device or the wearable and evaluate WearID's capability to differentiate people's voices when fixing the speech content. In particular, each participant's accelerometer data is compared with other participants' microphone data to calculate speech similarity. Figure~\ref{fig:conf_horizontal} shows the confusion matrix to differentiate people's voices between the VA device's microphone and the two smartwatches when they are held horizontally. We observe that WearID can accurately detect voice sounds received by the microphone and accelerometer to be from the same or different people.
In particular, Huawei Watch 2 shows an average of $96\%$ and the LG Urban W150 achieves $86\%$ accuracy.
Figure~\ref{fig:conf_vertical} further confirms our observation by showing the results of differentiating people's voice across two domains when the two smartwatches are held vertically. Specifically, Huawei Watch 2 obtain $91\%$ average accuracy while LG Urban W150 achieves $94\%$ accuracy. The results indicate that even under such extreme cases when people speak the same command, WearID can distinguish them correctly. 


\vspace{-2mm}
\subsubsection{Against Random Attack} Under the random attack, an adversary does not have prior information of the user's voice sound and try to use his/her own voice to bypass the VA system. Since the user is absent from the VA system, the voice sound received by the user's wearable (e.g., the user's voice sound) would be different from that recorded by the VA system. To evaluate the performance of WearID under random voice attacks, we let each participant alternatively performs as the legitimate user and be attacked by other participants. The wearable records the user's voice sound and the VA system's microphone records the adversaries' sound. Figure~\ref{fig:huawei2_roc2} and Figure~\ref{fig:w150_roc2} show the average ROC curves of WearID (e.g., blue) to verify the users with two different wearables under the two typical holding ways. We observe that WearID can verify the user and reject random attacks with high accuracy. In particular, the AUCs for Huawei watch 2 and LG Urbane W150 are $94.46\%$ and $88.85\%$ under the horizontal holding way. In addition, the vertical holding way shows slightly higher AUC, which are $96.81\%$ and $91.34\%$ for Huawei watch 2 and LG Urbane W150 respectively. Moreover, given a FPR of $5\%$, WearID can achieve high TPRs of $95.21\%$ and $98.47\%$ for Huawei watch 2 held in horizontal and vertical directions respectively.
The results indicate that WearID is effective to protect the VA system and verify the users with high accuracy under random attack. Moreover, in the practical scenarios, a legitimate user does not always speak and the wearable device usually records environmental noises. Thus the performance under random attack would approach to the normal situation (i.e., red curves).

\vspace{-1mm}
\subsubsection{Against Impersonation and Replay/Synthesis Attack}
We now consider the more sophisticated attacks on user's absence, which imitate/synthesize or just replay the legitimate user's voice commands to break the VA system. Ideally, an adversary could generate the voice sound which is exactly the same as the legitimate user. But the wearable is associated with the absent user and out of the adversary's control and it seldom happens when the two non-collocated devices (i.e., VA device and the wearable) receive the exactly same voice sounds from two independent sources. Because the wearable may still be possible to record the user's voice but with other speech content, we evaluate WearID in a more challenging scenario where the legitimate user's voice sounds are directly used for the attacking sounds of impersonation and replay attacks. 

Figure~\ref{fig:huawei2_roc2} and Figure~\ref{fig:w150_roc2} show the average ROC curve (i.e., black curves) when verifying the user under impersonation/replay attacks. We find that WearID successfully verify the user by using both Huawei watch 2 and LG Urbane W150 under both horizontal and vertical holding ways. In particular, WearID achieves $89.12\%$ and $86.78\%$ for Huawei Watch 2 and LG Urbane W150 under horizontal holding way. The AUCs are $91.23\%$ and $88.34\%$ under vertical holding way. For a FPR of $10\%$, WearID can obtain the TPRs of $91.25\%$ and $93.29\%$ when Huawei watch 2 is held in horizontal and vertical directions. We find the performance of WearID under impersonation and replay attacks are slightly lower than those obtained under random attacks. This is because the adversary has obtained the additional knowledge about the user's voice to improve the attack. But WearID still protects the user's privacy with high verification accuracy under this more challenging attacks. 
Moreover, in the practical scenarios, a legitimate user's wearable device does not always records the user's voice sounds, which make the performance approaching to that under normal situation.

\begin{figure}[t]
	\centering
	\includegraphics[width=2.1in]{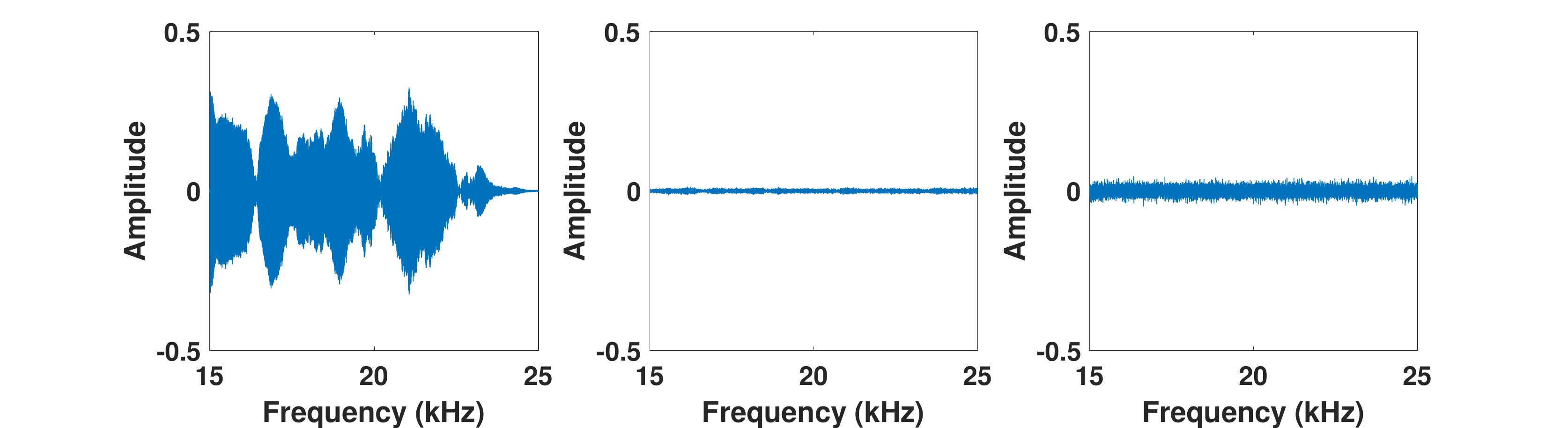}
	\vspace{-4mm}
	\caption{The frequency responses of the VA system and the wearables under ultrasound attacks.}
	\label{fig:ultrasound}
	\vspace{-7mm}
\end{figure}

\vspace{-1mm}
\subsection{User Verification under Co-location Attack}
Besides launching attacks when the user's absent, an adversary can also execute some special attacking sounds to both the VA devce and the user's wearable
without causing any notice. 

\begin{figure}[t]
	\begin{center}
		\begin{tabular}{cc}
			\includegraphics[width=1.5in]{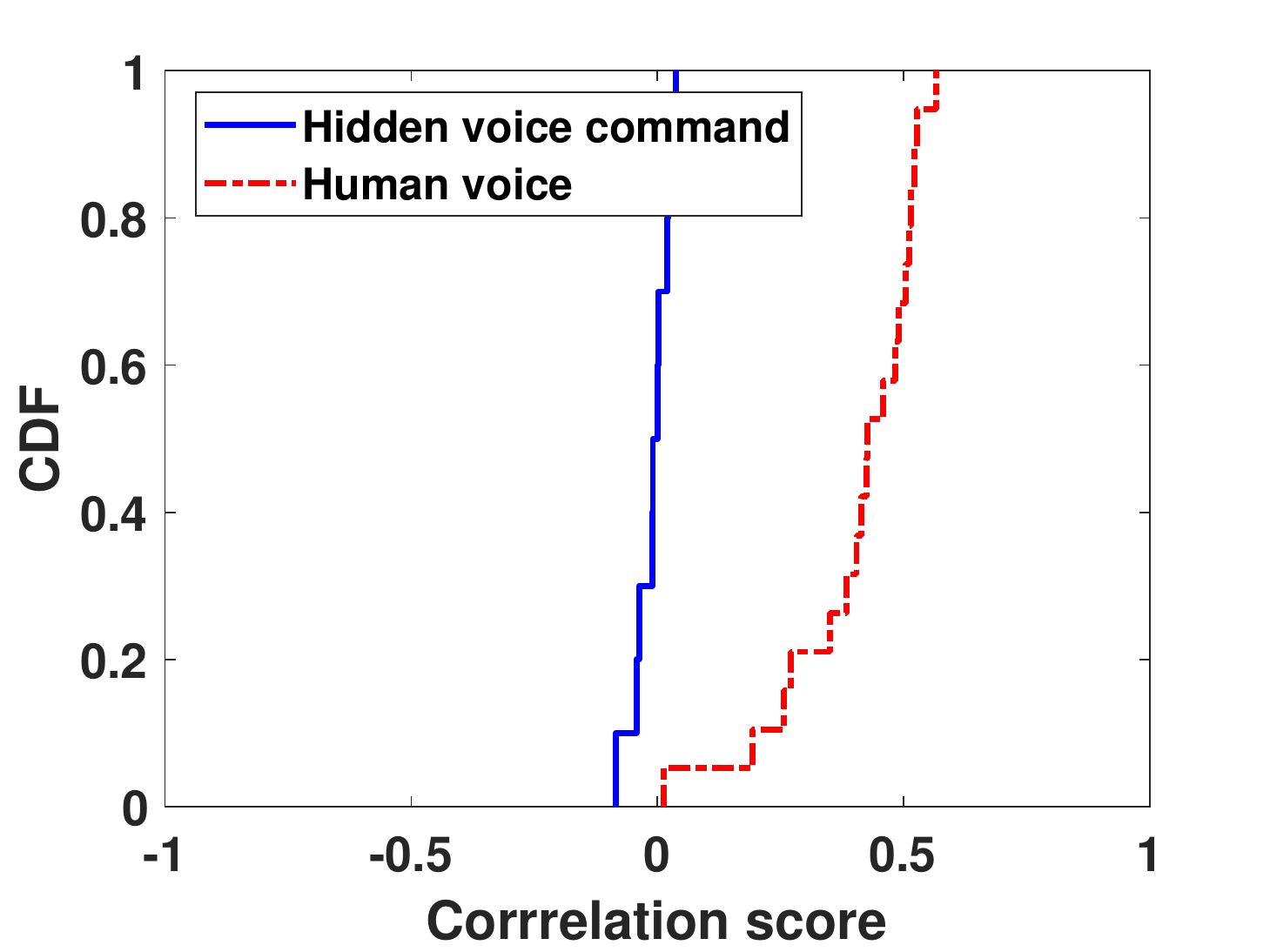}
			&
			\includegraphics[width=1.5in]{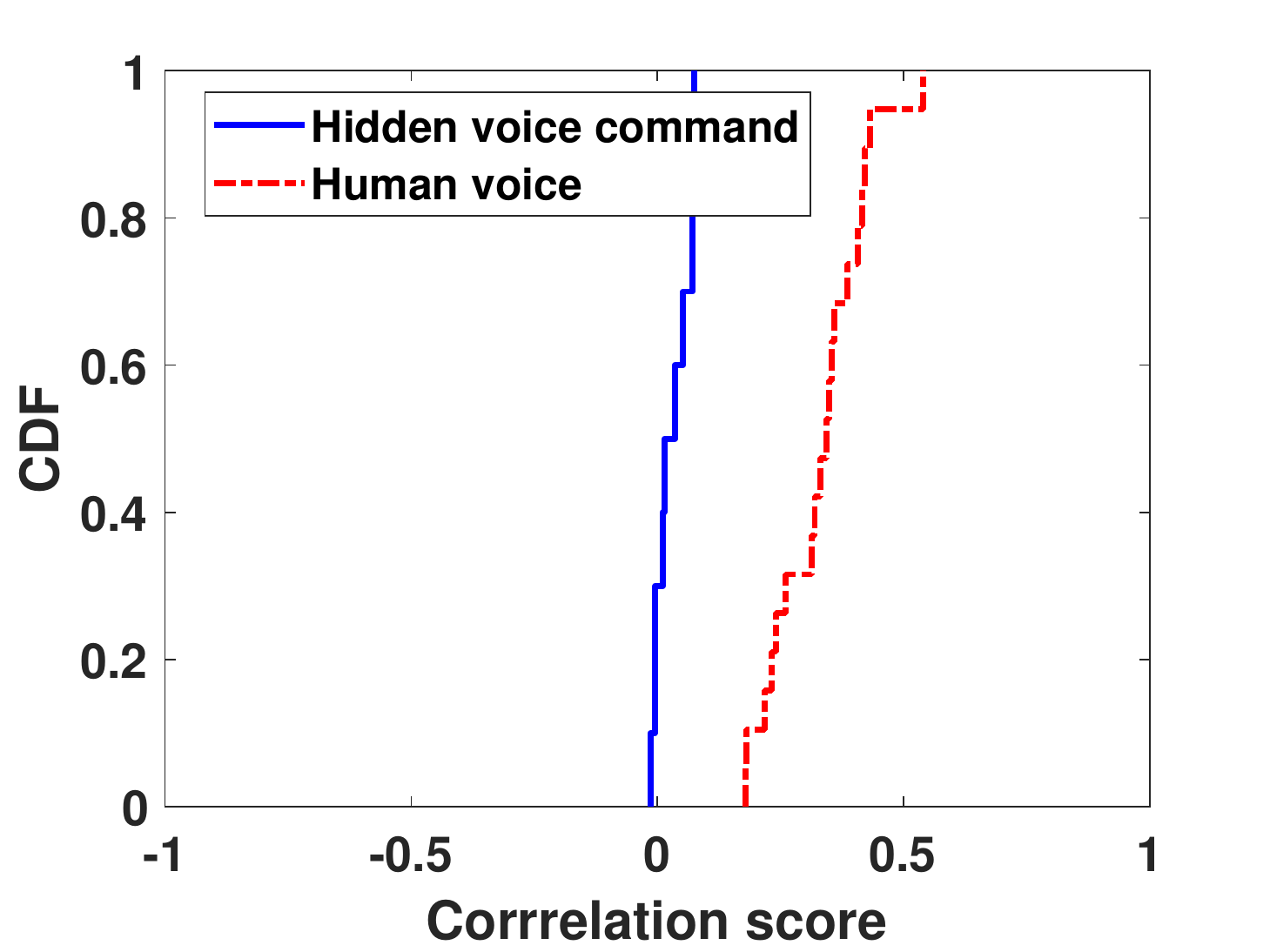}
			\\{\scriptsize(a) Huawei watch 2} & {\scriptsize(b) LG Urban W150}
		\end{tabular}
	\end{center}
	\vspace{-4mm}
	\caption{CDF of the 2D correlations between the microphone and the motion sensor data for the hidden voice commands and the legitimate user's voice commands.}
	\vspace{-7mm}
	\label{fig:hidden}
\end{figure}

\textbf{Against Hidden Voice Command.}
Under hidden voice attack, an adversary hides the recorded user voice sound into the noise sound, which is unintelligible to human but can be interpreted as commands by the VA system devices~\cite{carlini2016hidden}. The adversary then plays back such noisy sound using a loudspeaker to control the VA system without causing the user's notice. In such scenario, both the VA system's microphone and the user's wearable receive the hidden commands.
Figure~\ref{fig:hidden} depicts the CDFs of the 2D-correlations between the microphone data and motion sensor data under hidden voice commands, where the loudspeaker is placed $25$cm away to the two wearables and the volume is set to the maximum. We observe that the 2D-correlations between microphone and motion sensor are low for the hidden voice commands, which can be differentiated well from the legitimate user's voice commands. In particular, the median of the 2D-correlation coefficients for the hidden voice commands is around $0$ for Huawei watch 2 and $0.05$ for LG Urban W150. In comparison, the median 2D-correlation coefficients for the legitimate user's voice commands are around 0.5 for Huawei watch 2 and $0.4$ for LG Urban W150. The reason is that motion sensors on the wearable has short response distance and unique response characteristics to sound. An adversary is hard to fool the system which is developed based on verifying the sound from two domain information. The hidden voice attacks can thus be defended.

\textbf{Against Ultrasound Attack.}
Under the ultrasound attack, an adversary modulates the recorded user voice command to an inaudible frequency and plays back it using an ultrasound speaker. Such inaudible sounds can be recognized by the VA system but is hardly heard by the user~\cite{zhang2017dolphinattack}. In this scenario, both the VA's microphone and the user's wearable device is exposed to this inaudible sound. We thus evaluate WearID to see whether the ultrasound could leave similar responses on the both devices. In particular, we use a function generator (i.e., Keysight Technologies 33509B~\cite{Keysight33509B}) to generate a chirp of $15kHz\sim25kHz$ and play the chirp using a tweeter speaker~\cite{tweeterspeaker}, which is placed $25cm$ away from the wearable. Figure~\ref{fig:ultrasound} shows the frequency responses of VA microphone and the two smartwatches' accelerometers. We can find that the microphone show responses from $15kHz~24kHz$, which is a hardly heard frequency range. In comparison, we do not observe any responses on the two smartwatches. The experimental results that show even if the microphone is passed by the ultrasound attacks, the wearable's motion sensors reject the ultrasound.

	\vspace{-2mm}
\section{Discussion \& Conclusion}
\label{sec:conclusion}
In this paper, we present WearID, a wearable-assisted verification system for Voice Assistant (VA) systems (e.g., Amazon Echo and Google Home). WearID verifies whether the voice command received by the VA system comes from the legitimate user based on examining the command sound recorded in two domains (i.e., audio and vibration). In particular, WearID compares the voice command recorded by the VA device's microphone with that of the legitimate user's wearable motion sensor to calculate the cross-domain speech similarity. We show that the motion sensors of the wearable have a short response distance to sounds and exhibit different response characteristics from microphones. We further identify their complex relationship, which is hard to forge in various audible and inaudible acoustic attacks such as replay attacks and ultrasound attacks.
To compare the similarity between the two different sensing modalities under a huge sampling rate gap (e.g., 8000Hz vs. 200Hz), WearID converts the microphone data to low frequency data based on the unique frequency response characteristics of the motion sensor. The converted microphone data shows the similar spectrogram as that of the real motion sensor data. WearID then calculates shift 2D correlation between the spectrograms of the two data to verify the command sound. Moreover, WearID is easy to be deployed on the off-the-shelf wearable devices and does not require any hardware changes to the VA systems. Extensive experiments with two commodity smartwatches and 1000 commands are conducted and WearID is evaluated under both the attack on user's absence (e.g., replay attack) and the co-location attack (i.e., the hidden voice and the ultrasound attack). Experiments show that WearID can verify the command sound with 99.8\% accuracy in normal situation and detect 97\% fake voice commands under various acoustic attacks.

WearID is designed for the mass population of wearable users and is more secure than the existing methods such as virtual button ~\cite{virtualbutton2018} on the wearable devices, proximity detection (e.g., using Bluetooth and WiFi) and liveness detection~\cite{chen2017you}. All the three approaches have no or limited capability to verify the sound source and still suffer from many acoustic attacks such as hidden command attack and ultrasound attack. In addition, WearID requires minimum user effort, while the liveness detection requires user's mouth to be very close to the VA device and using a virtual button to access the VA system is quite cumbersome, which requires the user to unlock the wearable, find the App (or button), press the button and wait.



	
	\bibliographystyle{ACM-Reference-Format}
	\bibliography{./bib/main}
	
	\pagebreak
	\appendix
\section{Appendix}
\setcounter{figure}{0}   
\renewcommand{\thefigure}{A}
\begin{figure}[H]
	\centering
	\subfigure[Correlation matrix]{
		\includegraphics[width=2.2in]{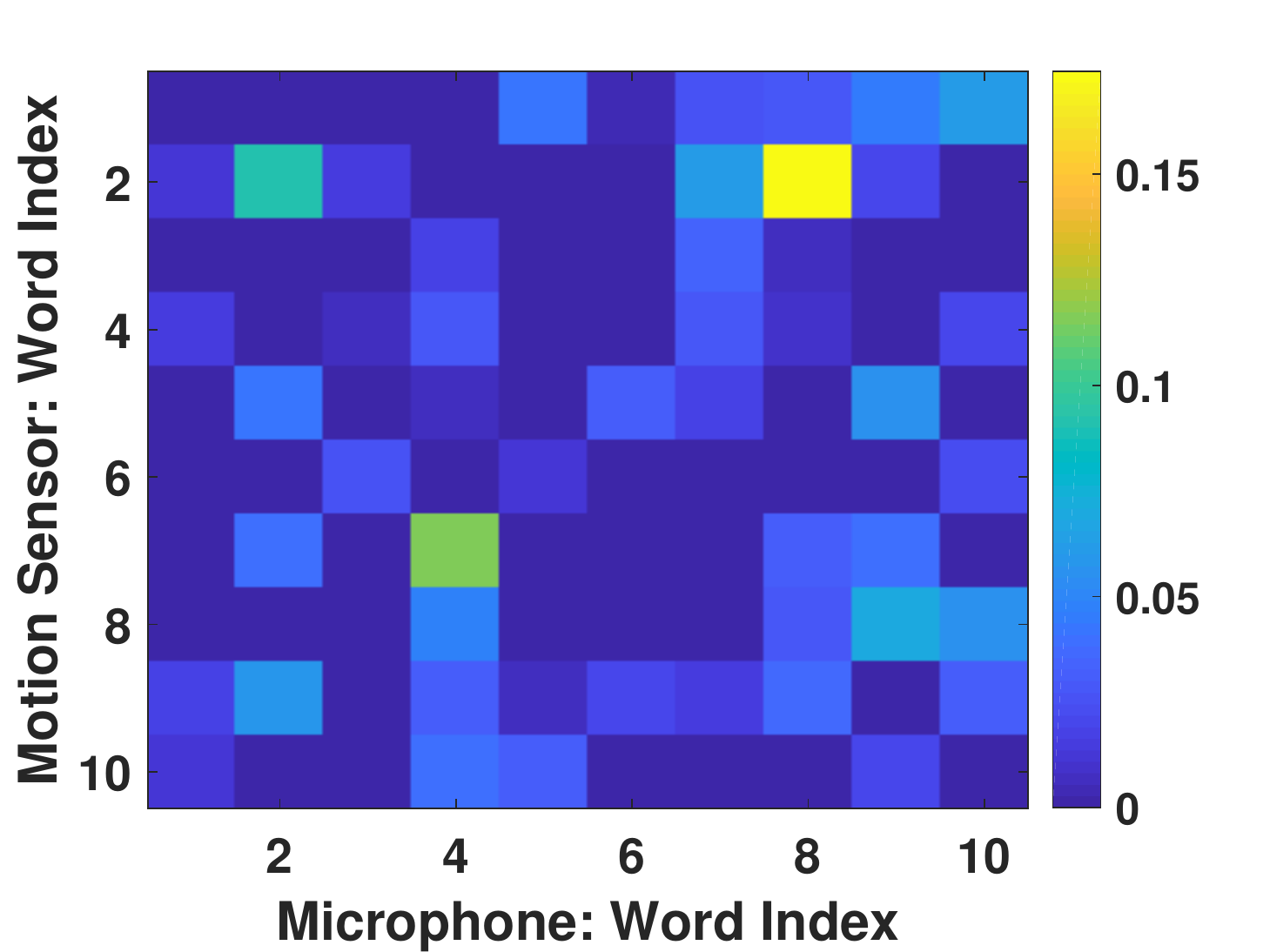}}
	\subfigure[CDF of the correlation]{
		\includegraphics[width=2.2in]{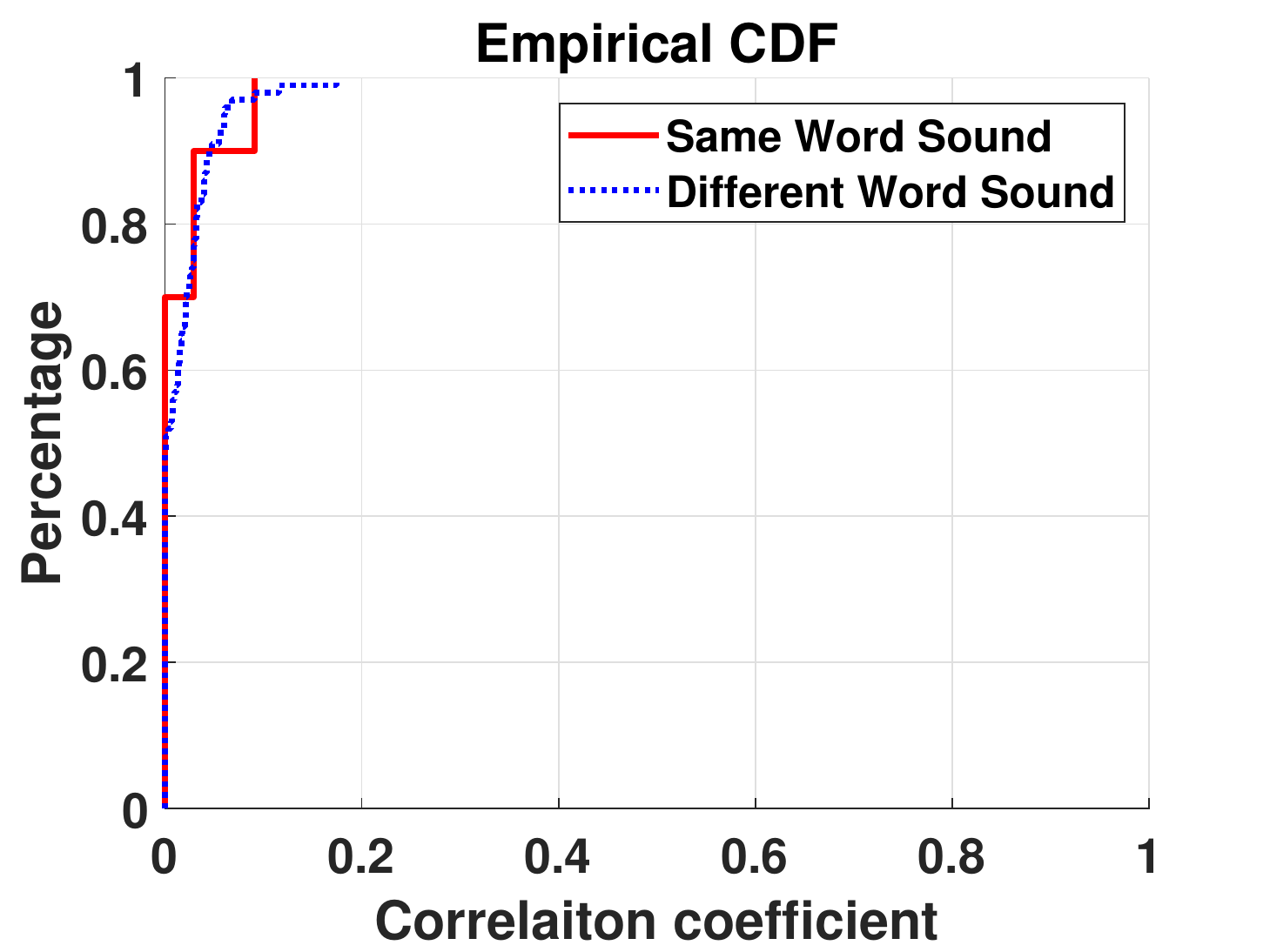}}
	\caption{The time-domain correlation between the microphone data and motion sensor, which are resampled to the same sampling rate level (Illustrated with 10 words on Amazon Echo and Huawei watch 2).}
	\label{fig:downsample_timecorrelation}
\end{figure}

\subsection{Difficulty of Comparing Microphone Data with Motion Sensor Data}
\label{subsec:difficulty}
Figure~\ref{fig:downsample_timecorrelation} illustrates the difficulty of comparing the microphone data with the motion sensor data, where a participant speaks ten words to both a microphone and a accelerometer, and both data are re-sampled to the same sampling rate for similarity comparison.
Particularly, Figure~\ref{fig:downsample_timecorrelation} (a) shows the time-domain correlation coefficient between the microphone recorded sound (i.e., X axis) and motion sensor data (i.e., Y axis) by cross-comparing ten words. We observe that the correlations at the diagonal (i.e., same word sound) and non-diagonal (i.e., different word sounds) are indistinguishable. The results indicate that the re-sampling technique and the time-domain analysis are insufficient to address the similarity comparison of the two different sensing modalities. Figure~\ref{fig:downsample_timecorrelation}(b), CDF of the correlation coefficients, further depicts the challenge of matching the sound across the two domains, where the sound of the same word and those of different words all show low correlation values (i.e., less than $0.1$). Thus, we need to investigate the inherent unique relationship between the two sensing modalities to facilitate their similarity comparison. 

\renewcommand{\thefigure}{B}
\begin{figure}[H]
	\centering
    \subfigure[Spectrogram of microphone]{
			\includegraphics[width=2.2in]{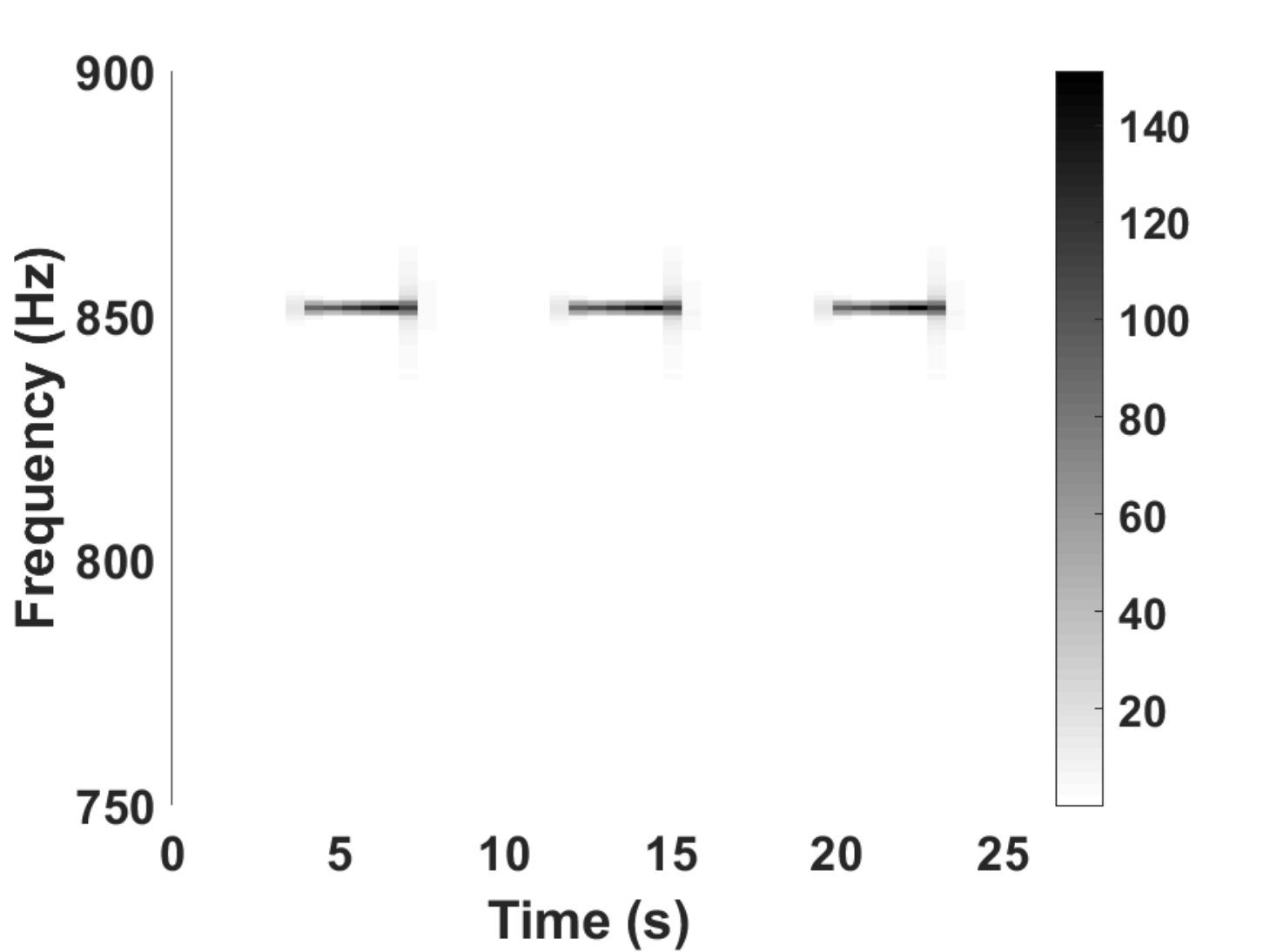}}
	\subfigure[Spectrogram of wearable accelerometer]{
			\includegraphics[width=2.2in]{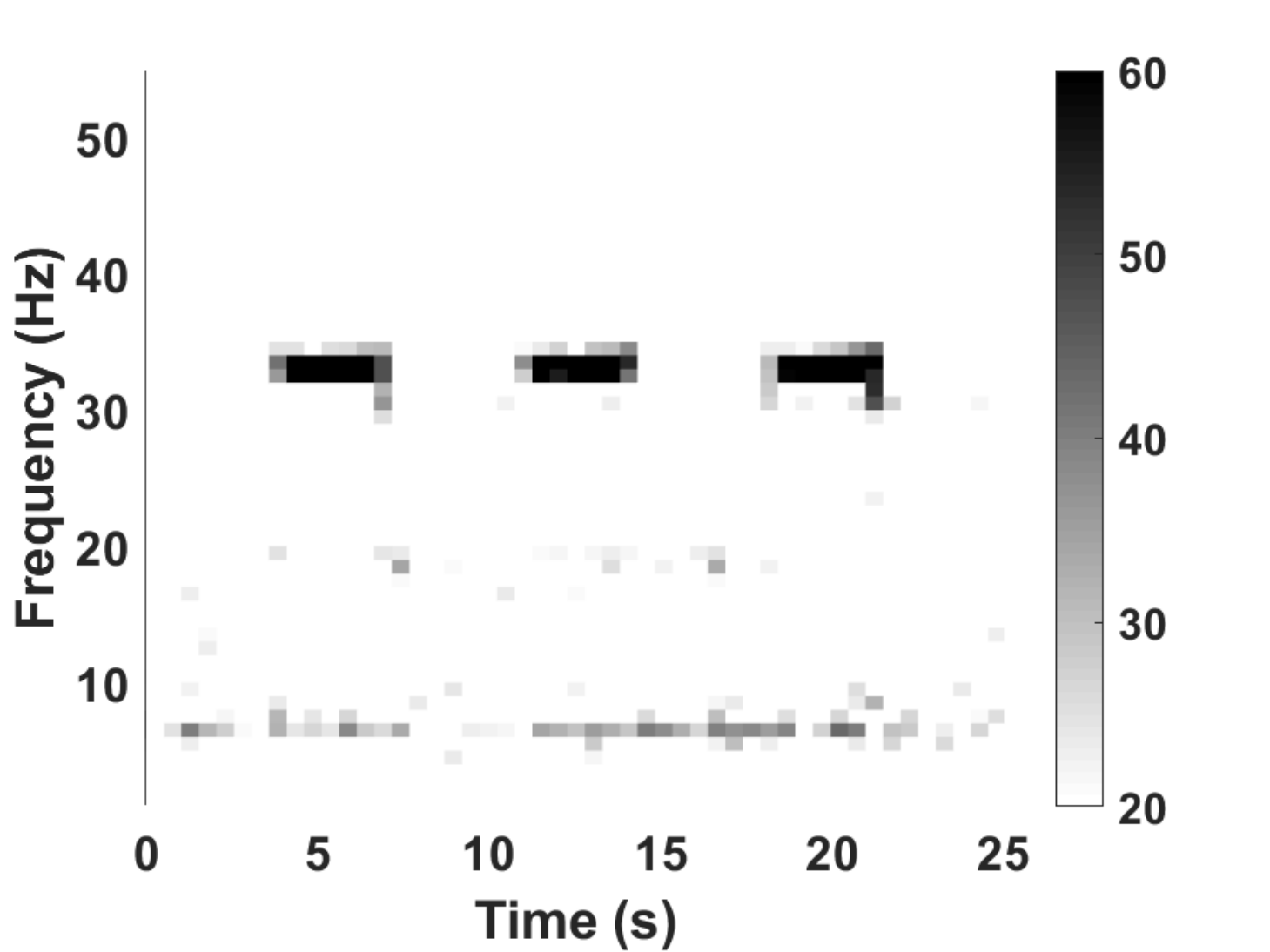}} 
	\caption{Spectrogram of the single frequency signal (850Hz) on microphone and wearable device (i.e., Huawei watch 2 sport).} 
	\label{fig:nonlinearity_singlefrequency}
\end{figure}

\subsection{Algorithm for Spectrogram based Conversion}
\begin{algorithm}[h]
	\footnotesize
	\caption{Spectrogram-based Conversion Algorithm}
	\begin{algorithmic}[2]
		\Function{Conversion}{$S_{mic}$}
		\State {\textbf{Input:} $S_{mic}$-original microphone spectrogram}
		\State {\ \ \ \ \ \ \ \ $f_{ws}$-sampling rate of accelerometer}
		\State {\textbf{Output:} $\hat{S_{mic}}$-converted microphone spectrogram}
		\State {$|\hat{S_{mic}}=zeros(T, F)|, \omega_{ws}=2\pi \times f_{ws}$}
		\For{$t=1:T$}
		\For{$f_mic=700:3300$} \State{// Frequency selection}
		\For{$N_{shift}=-10:10$}
		\State{$f_{w}=|f_{mic}-N_{shift}\times{f_{ws}}|$}
		\If{$|S_{mic}(t_n,f_m)|> 70dB \; \& \; f_{w}\leq{f_s} \; \& \; f_{w}>0$} \State{// Amplitude selection}
		\State {$\hat{S_{mic}}(t_n,f_w)=\hat{S_{mic}}(t_n,f_w)+|S_{mic}(t_n,f_m)|$} \State{// Spectrogram-based frequency conversion}
		\EndIf
		\EndFor
		\EndFor
		\EndFor
		\EndFunction
	\end{algorithmic}
	\label{al:spectrogram_transformation}
\end{algorithm}

\newpage 
\onecolumn
\subsection{Wearable Device Specifications}  
\setcounter{table}{0}
\renewcommand{\thetable}{A}
\begin{table*}[ht] 
    \begin{center}
    \caption{The specifications of the accelerometers in the tested wearable devices.}
    \label{table:devices}
    \begin{tabular}{| c | c | c | c | c | c|}
    \hline
    \scriptsize \textbf{Manufactory} & \scriptsize \textbf{Model} & \scriptsize \textbf{Accelerometer} & \scriptsize \textbf{User Programmable Range} &\scriptsize \textbf{Sensor sampling rate} &\scriptsize \textbf{System sampling rate} \\ \hline
      \scriptsize LG & \scriptsize Urbane watch 150 & \scriptsize Invensense M6515 & \scriptsize $\pm2g, \pm4g, \pm8g, \pm16g$ & \scriptsize 4-4000Hz & \scriptsize 200Hz  \\ \hline
      \scriptsize Huawei & \scriptsize Huawei watch 2 sport & \scriptsize STMicroelectronics LSM6DS3 & \scriptsize $\pm2g, \pm4g, \pm8g, \pm16g$ & \scriptsize 4-1600Hz & \scriptsize 100Hz  \\ \hline
    \end{tabular}
    \end{center}
\end{table*} 
\FloatBarrier
\subsection{Examples of the Voice Commands} 

\renewcommand{\thetable}{B}
\begin{table*}[ht]
\begin{center}
\caption{Example of privacy leakages from voice assistant systems.}
\label{tb:privacyleakage}
\begin{tabular}{|c|c|l|c|}
\hline
\scriptsize \textbf{Security issues} & \scriptsize \textbf{Category} & \scriptsize \textbf{Voice Command Examples} & \scriptsize \textbf{Sentence Length}\\ \hline
\multirow{11}{*}{\scriptsize \textbf{Potential privacy leakage}} & \multirow{4}{*}{\scriptsize \scriptsize \textbf{Event schedule} } & \scriptsize "What's on my calendar for tomorrow" & \scriptsize 6 \\  \cline{3-4}
                   &                   & \scriptsize "Where is my next appointment" & \scriptsize 5
 \\ \cline{3-4}
                   &                   & \scriptsize "List all events for January 1st" & \scriptsize 6
 \\ \cline{3-4}
                   &                   &  \scriptsize "How much is a round-trip flight to New York" & \scriptsize 9
 \\ \cline{2-4}
                   & \multirow{3}{*}{\scriptsize \textbf{Reminder}} & \scriptsize "Remember that my password is 'money'" & \scriptsize 6
 \\ \cline{3-4}
                   &                   & \scriptsize "What is my password" & \scriptsize 4
 \\ \cline{3-4}
                   &                   & \scriptsize "Add 'go to the grocery store' to my to-do list" & \scriptsize 10
 \\ \cline{2-4}
                   & \multirow{2}{*}{\scriptsize \textbf{Shopping account information}} & \scriptsize "What's on my shopping list" & \scriptsize 5 \\ \cline{3-4}
                   &                   &  \scriptsize "Track my order" & \scriptsize 3 \\ \cline{2-4}
                   & \multirow{2}{*}{\scriptsize \textbf{Contact}} & \scriptsize "Read me my email" & \scriptsize 4 \\ \cline{3-4}
                   &                             &  \scriptsize "Call my mother" & \scriptsize 3 \\ \hline
\multirow{9}{*}{\scriptsize \textbf{Unauthorized operation}}  & \multirow{2}{*}{\scriptsize \textbf{Neighborhood location}} & \scriptsize "Find me a Italian near my home" & \scriptsize 7 \\ \cline{3-4}
                   &                   & \scriptsize "What is the traffic to my home" & \scriptsize 7 \\ \cline{2-4}
                   & \multirow{2}{*}{\scriptsize \textbf{Unauthorized purchase}} & \scriptsize "Add paper towels to my cart" & \scriptsize 6 \\ \cline{3-4}
                   &                   & \scriptsize "Order all items in my cart" & \scriptsize 6 \\ \cline{2-4}
                   & \multirow{3}{*}{\scriptsize \textbf{Voice assistant}} & \scriptsize "Answer the call" & \scriptsize 3 \\ \cline{3-4}
                   &                   & \scriptsize "Delete all my reminders" & \scriptsize 4 \\ \cline{3-4}
                   &                   & \scriptsize "Play my favorite music on Spotify" & \scriptsize 6 \\ \cline{2-4}
                   & \multirow{2}{*}{\scriptsize \textbf{Access smart home devices}} & \scriptsize "Show the living room camera" & \scriptsize 5 \\ \cline{3-4}
                   &                   & \scriptsize "Clear all Bluetooth devices" & \scriptsize 4 \\ \hline
\end{tabular}
\end{center}
\end{table*}


\end{document}